\documentclass[twocolumn]{aastex63}

\newcommand{\be}{\begin{equation}}
\newcommand{\ee}{\end{equation}}

\usepackage{makecell}
\usepackage{amsmath}
\usepackage{amsfonts}
\usepackage{hyperref}
\usepackage[T1]{sansmath}
\usepackage{graphicx}

\begin{document}
\title{Dispersion Measure Variability in Fast Radio Bursts from Photoionization}

\author[0000-0002-4670-7509]{Brian D. Metzger}
\affiliation{Department of Physics and Columbia Astrophysics Laboratory, Columbia University, New York, NY 10027, USA}
\affil{Center for Computational Astrophysics, Flatiron Institute, 162 5th Ave, New York, NY 10010, USA}

\begin{abstract}
Magnetars became favored engines of fast radio bursts (FRBs) following the
discovery of a luminous radio burst coincident with a hard X-ray flare from a
Galactic magnetar. Several repeating FRB sources exhibit time-variable rotation
measures and compact, spatially coincident persistent radio emission, consistent
with energetic-particle nebulae confined by young supernova ejecta. Secular
changes in the dispersion measure (DM) of repeating FRBs have also been
observed, offering a complementary probe of their local environments; for
example, the DM of FRB 121102 rose until 2019 before declining in recent years.
Although rising DM evolution has been attributed to shock ionization, shocked
ejecta can cool efficiently through metal-line emission and recombine,
especially if mixed with cooler gas. Here we argue that DM variations of the
observed magnitude and timescale instead arise from changes in the ionization
state of supernova ejecta irradiated by X-rays from a time-variable central
engine. The dominant rapidly variable contribution comes from the dense, weakly
ionized shell swept up by the expanding nebula, whose ionization and
recombination times are shorter than those of the more extended ejecta. The
model predicts that enhanced FRB activity should be accompanied by rising DM,
with a response smoothed over the ionization/recombination time and superposed
on a slower secular decline from ejecta expansion. In FRB 121102, the DM maximum
occurred close to the burst-rich 2018--2019 activity episodes. If the recently
reported renewed activity is sustained, its DM decline should flatten and may
reverse into a fresh rise over the coming years.
\end{abstract}

\section{Introduction}

Fast radio bursts (FRBs) are luminous millisecond-duration radio transients
whose large dispersion measures (DMs) imply extragalactic distances
\citep{Lorimer2007,Thornton2013,CordesChatterjee2019,Petroff2022}. Even as the
physical mechanism responsible for the coherent radio emission remains
debated, magnetars have emerged as a leading central-engine candidate \citep{Popov2010,Lyubarsky2014}. This
interpretation was motivated early on by the discovery of the first repeating
source, FRB~121102, whose localization to a low-metallicity dwarf galaxy and
association with a compact persistent radio source (PRS) suggested a young, energetic
magnetar embedded within dense supernova ejecta
\citep{Chatterjee2017,Tendulkar2017,Marcote2017,Metzger2017,
Beloborodov2017,LuKumar2018}. It was later placed on firmer empirical footing by the
detection of an FRB-like radio burst from the Galactic magnetar
SGR~1935+2154, coincident with a hard X-ray flare
\citep{CHIMEFRB2020,Bochenek2020,Mereghetti2020,Tavani2021}.

The Galactic event also demonstrated that the radio burst can represent only a
small fraction of the total flare energy. For SGR~1935+2154, the inferred
X-ray-to-radio fluence ratio was
$\eta_{\rm X/r}\equiv {\cal F}_{\rm X}/{\cal F}_{\rm r}\sim 10^{5}$,
showing that the coherent radio emission is only the ``tip of the iceberg'' of
the engine's energy budget \citep{Bochenek2020,Li2021,Margalit2020}.
Although FRB~200428 was weak compared to bursts from cosmological repeaters,
the most energetic bursts from active repeaters can reach isotropic-equivalent
radio energies $\gtrsim10^{40}$--$10^{41}\,{\rm erg}$. FRB~121102 produces such luminous bursts during active epochs, with an energetic burst rate exceeding once per day
\citep{Law2017,Gourdji2019,Jahns2023,Li2021Nature,OuldBoukattine2024}. If these radio bursts are accompanied by X-ray-to-radio
energy ratios comparable to that of SGR~1935+2154, the
associated high-energy flare output can approach the
$\sim10^{44}-10^{46}\,{\rm erg}$ scale of Galactic magnetar giant flares
\citep{Hurley1999,Feroci2001,Palmer2005,Hurley2005}. Searches for simultaneous X-ray counterparts to extragalactic FRBs have so far
yielded nondetections, but the resulting limits remain broadly consistent with
large X-ray-to-radio ratios. For example, observations of the nearby active
repeater FRB~20240114A constrain
$\eta_{\rm X/r}\lesssim2.4\times10^{6}$ for an SGR~1935+2154-like spectrum
\citep{Eppel2025}.

FRB~121102 remains the clearest example of an FRB embedded in an extreme local
environment. In addition to its compact PRS, it
exhibits an enormous and time-variable rotation measure (RM), initially of
order $10^{5}\,{\rm rad\,m^{-2}}$ \citep{Michilli2018}. Both the persistent
synchrotron emission and the large RM can be explained by a compact,
magnetized electron-ion nebula inflated by repeated magnetar outflows and
confined by young supernova ejecta \citep{MargalitMetzger2018}. The physical
size of this nebula is relatively well constrained: synchrotron
self-absorption limits require it not to be too compact \citep{MargalitMetzger2018}, while VLBI
non-resolution of the PRS requires a diameter $\lesssim0.7$ pc
\citep{Marcote2017,Snelders2025}. Together these arguments motivate a
characteristic nebular radius
$0.03\,{\rm pc}\lesssim R_{\rm n}\lesssim0.3\,{\rm pc}$. 

A crucial feature of this picture is that the nebula is not an ordinary
rotation-powered pulsar wind nebula composed primarily of electron-positron
pairs; pairs contribute little to the cold-plasma DM and produce no net
Faraday rotation. Instead, the RM and persistent synchrotron emission point to
a baryon-loaded, magnetically powered nebula, plausibly fed by repeated
magnetar flares \citep{Beloborodov2017,MargalitMetzger2018,Metzger2019}.
The recent discovery of nuclear de-excitation gamma-ray lines from the 2004
giant flare of SGR~1806--20 further supports the idea that magnetar flares can
eject substantial baryonic material at trans-relativistic speeds into their
surroundings \citep{Cehula2024,Patel2025}. Fluctuations in RM are then naturally expected on the turbulent overturn time
of the nebula, which for trans-relativistic turbulence is of order
$R_{\rm n}/c\sim0.1$--$1\,{\rm yr}$, broadly consistent with the RM variability
observed in PRS-associated repeaters
\citep{Michilli2018,Hilmarsson2021,AnnaThomas2023,Moroianu2025}. Repeated
magnetar flares may also stir the nebular plasma and drive departures from a
smooth secular RM decline, suggesting that epochs of enhanced bursting activity
could be accompanied by increased RM variability.

PRSs are now known, or strongly suspected, around several other repeating
FRBs. Compact persistent counterparts have been reported for FRB~20190520B,
FRB~20240114A, and FRB~20190417A, in addition to FRB~121102
\citep{Niu2022,Bhandari2023,Bruni2025,Moroianu2025,Pelliciari2026}. These systems share some of the properties that make FRB~121102 distinctive,
including compact non-thermal radio emission, large or time-variable RMs, and,
in the best-localized cases, low-metallicity dwarf-galaxy hosts \citep{Law2022}. A still fainter candidate PRS has also been identified near FRB~20181030A,
potentially extending the FRB--PRS population to lower radio luminosities and
more modest RMs \citep{Ibik2024,Pelliciari2026}. Taken together, these sources
suggest that at least a subset of repeating FRBs are surrounded by dense,
magnetized, evolving plasma environments.

The DM provides a complementary probe of this local material. Unlike the RM,
which depends on the line-of-sight magnetic field and can vary because of
turbulence or changes in field geometry, the DM directly measures the column of
non-relativistic free electrons. The relativistic nebula responsible for the
PRS contributes little to the observed DM because its radiating particles are both too few in number and too relativistic to behave
as a cold dispersive plasma. A substantial local DM therefore
requires an additional, colder and denser reservoir of ionized plasma. In the
young magnetar scenario, the natural source of this material is the ionic ejecta
shell that confines the nebula.  Young supernova ejecta around FRB sources were therefore
identified early on as natural sources of evolving local DM
\citep{Connor2016,Piro2016,YangZhang2017,Metzger2017,PiroGaensler2018}.

Such secular evolution of the DM is now observed in several repeaters. FRB~121102 has
shown significant DM evolution over the past decade, including a rise to a
maximum around 2019 followed by a decline over the subsequent years
\citep{Hessels2019,Snelders2025,Wang2025,Waxman2026}. The long-baseline
FAST/GBT monitoring presented by \citet{Wang2025} strengthens the evidence for
a genuine post-2019 decrease in the source-local DM and shows that the large
RM has continued to decline, while also emphasizing that the radio burst
activity remains highly intermittent rather than monotonically fading.
FRB~20190520B also possesses a large source-local DM and RM, with evidence for
secular DM decrease \citep{Niu2022,ZhaoWang2021,AnnaThomas2023,Niu2026}. More
recently, FRB~20220529A has been reported to show a steady DM decline,
together with a short-lived DM/RM excursion suggestive of a dynamic local
magneto-ionic environment \citep{Pandhi2026}. Population-level studies of repeating FRBs also find evidence for secular DM evolution, although the inferred incidence depends on how burst structure and observational systematics are treated. In particular, \citet{Cook2026} find
significant DM variability in at least 7 of 81 CHIME/FRB repeaters using
higher-time-resolution data and accounting for the degeneracy between DM and
downward sub-burst frequency drift, while \citet{Cui2026} report a larger
candidate fraction based on real-time CHIME/FRB alert DMs.

In the simplest young-supernova models for local DM, the ejecta expands
homologously and the ionized fraction is prescribed or assumed to vary slowly,
so expansion of a fixed-mass shell gives
${\rm DM}\propto t^{-2}$ \citep[e.g.,][]{Connor2016,Metzger2017,
YangZhang2017,Cui2026}. More detailed supernova-remnant models instead tie the
ionized column to shock dynamics: \citet{Piro2016} and
\citet{PiroGaensler2018} calculated the DM and RM from shocked supernova ejecta
and circumstellar material, with the ionized ejecta mass controlled largely by
the propagation of the reverse shock. Along similar lines, \citet{Waxman2026} attributed the DM evolution of FRB~121102 to the expansion of a shocked shell driven by the PRS plasma into
cold surrounding ejecta. In their interpretation, the observed rate of DM
evolution constrains the confining material to expand slowly,
$v_0\lesssim 200\,{\rm km\,s^{-1}}$, with kinetic energy
$\sim 10^{47}\,{\rm erg}$ and swept-up mass of order
$0.3$--$3\,M_\odot$.

However, in either case, the observed DM should not necessarily be identified
with the total ejecta column or even the total shocked column. At the densities
expected near the nebula-ejecta interface, shocked gas can cool efficiently by
metal-line emission, compress into a dense shell, and recombine. Indeed,
models of young pulsar wind nebulae find that the relativistic bubble sweeps
freely expanding ejecta into a dense, radiatively cooled, Rayleigh-Taylor
unstable layer of filaments \citep{ChevalierFransson1992,Jun1998,Blondin2001},
broadly similar to the filamentary structures observed in young pulsar-wind nebulae such as
the Crab Nebula. Multidimensional instabilities and turbulent mixing with cooler material can
further enhance the loss of thermal energy \citep{Metzger2025}.\footnote{A
closely related effect is illustrated by classical novae, where internal shocks
occur as a fast white-dwarf outflow impacts a dense shell of gas released
earlier in the outburst, generating powerful gamma-ray emission
\citep{Chomiuk2021}. However, the observed X-ray luminosities of nova shocks
are suppressed by $\sim4$--$5$ orders of magnitude relative to one-dimensional
radiative shock-model predictions \citep[e.g.,][]{Nelson2019}, likely because
turbulent mixing of post-shock gas with previously swept-up cool
($\lesssim10^4\,{\rm K}$) gas greatly dominates direct radiative losses from the hot phase \citep{Steinberg2018,Metzger2025}.} Shock processing alone therefore may fail
to maintain a large ionized column, and could even reduce the free-electron
content by placing material into a colder, denser recombining layer.

\begin{figure*}
\centering
\includegraphics[width=\textwidth]{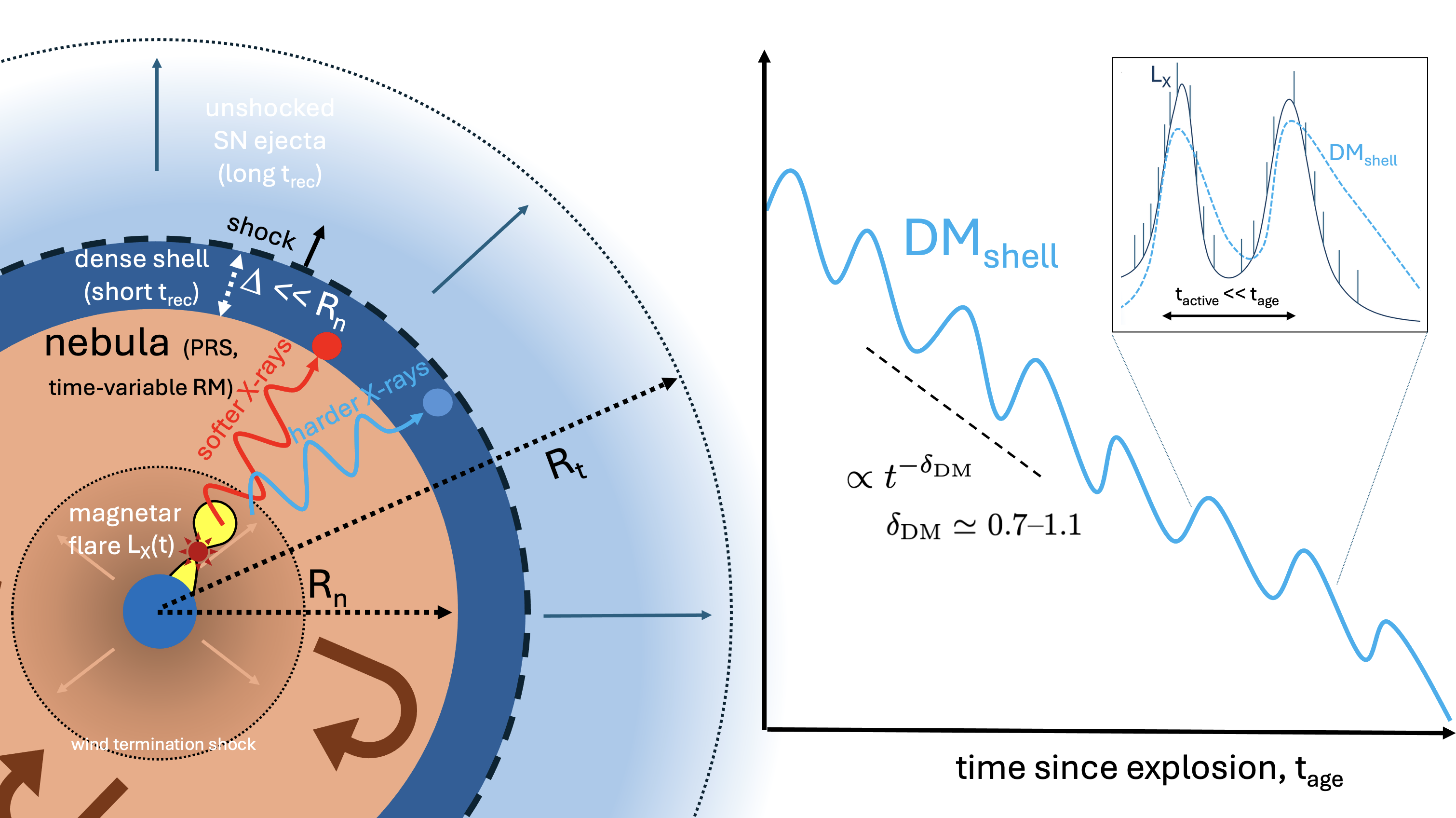}
\caption{
Schematic illustration of the photoionization-regulated DM scenario. A central magnetar
launches a relativistic wind that is thermalized at a wind termination shock
and inflates a magnetized nebula, identified with the persistent radio source
(PRS). Turbulence and magnetic-field fluctuations inside the nebula can give
rise to the large, time-variable RM, while its secular decline reflects the
expansion and dilution of the magnetized nebula confined by the surrounding
ejecta \citep{MargalitMetzger2018}. The nebula drives a forward shock into the
freely expanding supernova ejecta, sweeping the inner ejecta into a thin dense
shell of thickness $\Delta\ll R_{\rm n}$. This dense shell has a short
recombination time and is photoionized by the central X-ray luminosity:
softer X-rays are absorbed closer to the inner edge, while harder X-rays
penetrate to larger columns. The resulting ionized column produces the
time-variable shell contribution, ${\rm DM}_{\rm shell}$. More extended,
unswept ejecta near the transition radius $R_t$ have lower density and longer
recombination times, and therefore contribute mainly to a slowly varying or
approximately constant local DM. Right: on the system age, the shell DM
declines secularly as the ejecta expand, approximately
${\rm DM}_{\rm shell}\propto t_{\rm age}^{-\delta_{\rm DM}}$. On shorter
activity timescales, $t_{\rm active}\ll t_{\rm age}$, individual magnetar
flares and longer active episodes in $L_{\rm X}(t)$ drive a smoothed and
delayed DM response as the shell approaches and departs from
photoionization-recombination equilibrium.
}
\label{fig:cartoon}
\end{figure*}

The large output of high-energy radiation from magnetar flares suggests a
different possibility. As discussed above, the coherent radio burst likely
represents only a small fraction of the total flare energy, with the bulk
emerging at X-ray energies. This motivates the alternative
explored here: the time-variable DM is regulated primarily by photoionization
from the central engine. In our model, X-rays from repeated
magnetar flares penetrate into the inner regions of the supernova ejecta, where
the ionization fraction is set by a balance between photoionization and
radiative recombination. Changes in the time-averaged X-ray luminosity alter
this balance: during episodes of enhanced high-energy activity, the equilibrium
ionized column rises and the DM increases, while after the activity subsides
the ejecta recombine and the DM declines.
We explore whether this mechanism can account for the observed DM evolution of
FRB~121102. As we show below, an ionization-regulated local DM naturally gives both the observed magnitude, ${\rm DM}_{\rm local}\sim10$--$100\,{\rm pc\,cm^{-3}}$, and year-to-decade evolution timescales for ejecta masses, explosion energies, and long-term engine powers compatible with the standard picture of a young magnetar embedded
inside supernova ejecta.

The rest of this paper is organized as follows. In Section~\ref{sec:model} we
introduce a generic thin-shell model for the dense gas surrounding the FRB
source, calculate the X-ray photoionization rate as a function of depth, and
derive the resulting equilibrium ionized column, DM, and recombination
timescale. In Section~\ref{sec:time-dependent}, we calculate the time-dependent response
of the shell DM to finite episodes of enhanced X-ray activity. In Section~\ref{sec:discussion} we
apply these results to supernova ejecta shells and to FRB~121102, and discuss
the challenges to previous shock-ionization models once post-shock cooling and
recombination are included. In Section~\ref{sec:conclusions}, we discuss additional predictions of the
scenario and conclude. Appendices~\ref{sec:thermal-balance} and \ref{sec:sn-shell} supplement the
main text by discussing the thermal balance of photoionized gas in the swept-up
shell and reviewing the dynamics of young supernova ejecta swept up by a
magnetar wind nebula, respectively.

\section{Model}
\label{sec:model}

We begin with a phenomenological model for the cold material surrounding the
FRB source. Motivated by the picture described above, we take this material to
reside in a dense shell outside the wind nebula of the central engine. We focus
on this inner shell because its small radius and high density make it the most
natural site for a local DM component that can evolve on year-to-decade
timescales. Our fiducial interpretation is a magnetar wind nebula confined
within young supernova ejecta; the connection between the shell parameters and
the supernova-ejecta model is summarized in Section~\ref{sec:remnant} and
derived in Appendix~\ref{sec:sn-shell}. However, the photoionization calculation
itself is agnostic to the nature of the FRB engine and the origin of the dense
shell. Qualitatively similar conditions could also arise in other FRB models,
such as wind-inflated bubbles from mass-transferring X-ray binaries
\citep[e.g.,][]{Sridhar2021,Sridhar2022}. Figure~\ref{fig:cartoon} provides a
schematic illustration of the physical picture, which we develop below.

We consider a stationary thin shell of mass $M=M_1M_{\odot}$ at radius
$R=0.1R_{0.1}\,{\rm pc}$, roughly corresponding to the outer edge of the
nebula. The shell has radial thickness $\Delta=f_\Delta R$, with
$f_\Delta\ll1$, and uniform mass density
\be
\rho \simeq \frac{M}{4\pi R^{2}\Delta}
\simeq 5.4\times 10^{-19}\,{\rm g\,cm^{-3}}\,
M_{1}R_{0.1}^{-3}
\left(\frac{f_\Delta}{0.01}\right)^{-1}.
\ee
A geometrically thin shell with $f_{\Delta} \lesssim 10^{-2}$ is motivated by the efficient radiative cooling of
ejecta swept up by the expanding nebula (Appendix~\ref{sec:sn-shell}).
The total mass column through the shell is
\be
\Sigma_{\rm tot} \simeq \rho \Delta
= \frac{M}{4\pi R^{2}}
\simeq 1.7\times 10^{-3}\,{\rm g\,cm^{-2}}\,
M_1 R_{0.1}^{-2}.
\ee
We hereafter use $0 < \Sigma/\Sigma_{\rm tot} < 1$ as a coordinate through the shell.

For simplicity, we follow only the ionization balance of hydrogen and adopt a
fiducial hydrogen mass fraction $X=0.7$. Both assumptions are idealizations:
inner supernova ejecta can contain a substantial admixture of heavier elements,
and the hydrogen abundance near the nebula-ejecta interface need not be
solar-like. We nevertheless focus on hydrogen because, when present, it is
relatively easy to ionize completely and provides the largest number of free
electrons per unit mass. Although simplified, the estimates below can be
generalized straightforwardly to different compositions or to include metal
ionization, which we do not expect to change the qualitative conclusions.

If a fraction $x$ of the hydrogen atoms are ionized, the resulting electron density is
\be
n_e = \frac{\rho X x}{m_p}
\simeq 2.3\times 10^{5}\,{\rm cm^{-3}}\,
x\,
M_{1}R_{0.1}^{-3}
\left(\frac{f_\Delta}{0.01}\right)^{-1}.
\ee
The resulting dispersion measure interior to column $\Sigma$ is
\be
{\rm DM}(\Sigma) = n_e \Delta = \frac{X\Sigma x}{m_p} = {\rm DM}_{\rm max}
\left(\frac{\Sigma}{\Sigma_{\rm tot}}\right)x,
\ee
where ${\rm DM}_{\rm max}\equiv X\Sigma_{\rm tot}/m_p\simeq
220M_1R_{0.1}^{-2}\,{\rm pc\,cm^{-3}}$ is the DM of a fully
ionized shell.  We will find below that the shell is typically weakly ionized, $x\ll1$, so its
DM contribution is much smaller than ${\rm DM}_{\rm max}$ even though the fully
ionized value is comparable to the total DM of many FRBs.

\subsection{X-ray Emission from the Flaring FRB Engine}

We next estimate the ionization state produced by irradiation from a central
X-ray source of luminosity $L_{\rm X}$. Our fiducial picture is that this ionizing luminosity is dominated by the
time-averaged high-energy output of repeated flares from the FRB engine (e.g., magnetar). As discussed near the end of the paper, the DM
change produced by any individual flare is likely modest, except perhaps for
the most energetic events. The relevant quantity is therefore the ionizing luminosity averaged over the
time on which the ionized fraction changes appreciably, which below is set by
the local recombination time of the shell of typically years.

We assume a power-law photon spectrum
$d\dot{N}/dE\propto E^{-\Gamma}$ extending from $E_{\rm min}$ to $E_{\rm max}$.
As motivated by FRB~200428 from SGR~1935+2154, which was accompanied by a hard
X-ray counterpart detected by multiple high-energy instruments
\citep{CHIMEFRB2020,Bochenek2020,Mereghetti2020,Ridnaia2021,Li2021,Tavani2021},
we take $\Gamma\simeq0.7$ and $E_{\rm max}\simeq50\,{\rm keV}$ as fiducial
values in our estimate below, broadly based on the INTEGRAL spectral fit of \citet{Mereghetti2020}. More generally, magnetar bursts and giant
flares exhibit a broad range of spectral hardness, with characteristic photon
energies from tens of keV in ordinary short bursts to hundreds of keV or more
in the initial spikes of giant flares
\citep[e.g.,][]{Hurley1999,Palmer2005,Hurley2005,KaspiBeloborodov2017,Godwin2026}. We keep the dependence on $E_{\rm max}$ explicit below.

Under this assumption, the specific X-ray luminosity at the source (i.e., prior to any attenuation) can be written,
\begin{eqnarray}
\left(\frac{dL}{dE}\right)_0 &=&
\frac{(2-\Gamma)L_{\rm X}}{E_{\rm max}^{2-\Gamma}-E_{\rm min}^{2-\Gamma}}
E^{1-\Gamma} \nonumber \\
&\underset{E_{\rm max}\gg E_{\rm min}}\simeq& 
(2-\Gamma)\frac{L_{\rm X}}{E_{\rm max}}
\left(\frac{E}{E_{\rm max}}\right)^{1-\Gamma}.
\end{eqnarray}
We parametrize the time-averaged X-ray luminosity as a fraction $f_{\rm X}$ of the
long-term engine power,
\begin{eqnarray}
L_{\rm X}
&\sim&
f_{\rm X}\frac{E_{\rm mag}}{t_{\rm age}}
 \nonumber \\
&\simeq& 6.3\times10^{39}\,{\rm erg\,s^{-1}}\,
\left(\frac{f_{\rm X}}{0.1}\right)
E_{{\rm mag},50}
\left(\frac{t_{\rm age}}{50\,{\rm yr}}\right)^{-1},
\label{eq:LX_engine}
\end{eqnarray}
where $E_{\rm mag}$ is the total energy released by the engine over the source age
$t_{\rm age}$. The fiducial normalization is motivated by FRB~121102: modeling its persistent
synchrotron nebula, \citet{MargalitMetzger2018} inferred a total injected
energy $E_{\rm mag}\sim10^{50}$--$10^{51}\,{\rm erg}$ over an age $t_{\rm age}$ of decades
to a century. This energy scale may be broadly consistent with the magnetic energy
available from a young hyper-active magnetar \citep{Beloborodov2017}, albeit depending on details such as how efficiently strong fields diffuse out of the magnetar's core \citep{BeloborodovLi2016,Bransgrove2026}.  

A similar scale is obtained by estimating the time-averaged flare power
directly from the burst activity of FRB~121102. During active periods,
FRB~121102 produces luminous radio bursts with isotropic-equivalent energies
$E_{\rm r}\gtrsim10^{40}\,{\rm erg}$ at rates of order
$\mathcal{R}(>E_{\rm r})\sim0.03\,{\rm hr^{-1}}$
(\citealt{Law2017,Gourdji2019,Nicholl2017}; see also
\citealt{Margalit2020}). If the accompanying high-energy fluence exceeds the
radio fluence by a factor
$\eta_{\rm X/r}\equiv{\cal F}_{\rm X}/{\cal F}_{\rm r}\sim10^{5}$, comparable
to that inferred for SGR~1935+2154
\citep{Bochenek2020,Li2021,Margalit2020}, the implied time-averaged ionizing
luminosity would be
\begin{eqnarray}
L_{\rm X}
&\sim&
\eta_{\rm X/r}E_{\rm r}\mathcal{R} \nonumber \\
&\simeq&
8\times10^{39}\,{\rm erg\,s^{-1}}\,
\left(\frac{\eta_{\rm X/r}}{10^{5}}\right)
\left(\frac{E_{\rm r}}{10^{40}\,{\rm erg}}\right)
\left(\frac{\mathcal{R}}{0.03\,{\rm hr^{-1}}}\right), \nonumber \\
\label{eq:LX_bursts}
\end{eqnarray}
consistent with the independent estimate in Eq.~\eqref{eq:LX_engine}.

Even as a time-average, we should not expect $L_X$ to remain constant in time. Repeating FRBs are highly intermittent, exhibiting clustered burst activity,
long inactive intervals, possible activity cycles, and changes in the overall
activity level from one epoch to another. FRB~121102, for example, shows
strongly clustered activity with evidence for a $\simeq157$ day activity cycle
\citep{Gourdji2019,Cruces2021,Rajwade2020,Braga2025,Wang2025}, while
FRB~20180916B exhibits a well-defined $\simeq16.3$ day activity cycle
\citep{CHIMEFRB2020periodic}. Galactic magnetars likewise undergo episodic
active states in which both the bursting rate and persistent X-ray luminosity
increase and then decay over months to years
\citep[e.g.,][]{ReaEsposito2011,CotiZelati2018}. Variations in $L_{\rm X}$ on
timescales from activity cycles to longer-term active and quiescent epochs are
therefore a natural ingredient of the model and motivate the time-dependent
ionization calculations below.

The nebular forward shock may also contribute ionizing radiation. For the
parameters of interest, this shock is radiative
(Appendix~\ref{sec:sn-shell}), but its power is only a small fraction of the
magnetar injection power, $L_{\rm sh}/\dot{E}_{\rm mag}\sim
{\rm few}\times10^{-3}$, because it slowly overtakes the homologously expanding
inner ejecta, $v_{\rm sh}=dR_{\rm n}/dt-R_{\rm n}/t\simeq R_{\rm n}/5t$.
Thus, shock-powered ionization is likely subdominant if even a percent-level
fraction of the time-averaged magnetar power emerges as flare-powered ionizing
radiation, though it may provide a steadier ionizing floor during quiet epochs.

\subsection{Photo-ionization}
\label{sec:photoionization}

We next estimate how deeply the central X-ray radiation penetrates into the
shell and how much ionization it produces as a function of column. For
simplicity, we follow only the ionization balance of hydrogen, which dominates
the free-electron budget by number and therefore
captures the leading scaling of the DM with luminosity, density, and depth.
Metals are not followed as separate ion species, but their contribution is
included implicitly through the adopted photoelectric opacity law and through
the metal-line cooling rates used to estimate the gas temperature entering the recombination rate.

As the X-rays propagate into the shell, they are absorbed primarily through
bound-free transitions. For the order-of-magnitude estimates below, we approximate the effective bound-free cross section, normalized at the
hydrogen ionization threshold, as 
\be
\sigma_{\rm bf} \approx \sigma_{\rm H}
\left(\frac{E}{\chi_{\rm H}}\right)^{-\beta},
\ee
where $\chi_{\rm H}=13.6\,{\rm eV}$, $\sigma_{\rm H}=6.3\times10^{-18}\,{\rm cm^2}$,
and we take $\beta=2.5$ as a representative value for solar-composition gas
over the relevant soft X-ray band \citep[e.g.,][]{Cruddace1974}. The
corresponding mass absorption coefficient is
\be
\kappa_{\rm bf} =
\frac{\sigma_{\rm bf}X(1-x)}{m_p}
\underset{x\ll 1}\simeq
\kappa_{\rm H}
\left(\frac{E}{\chi_{\rm H}}\right)^{-\beta},
\ee
where $\kappa_{\rm H} \equiv \sigma_{\rm H} X/m_p
\simeq 2.6\times10^6\,{\rm cm^2\,g^{-1}}$ for $X = 0.7$ and we assume that the gas is
mostly neutral $x \ll 1$, an approximation checked below. The optical depth from the inner edge of the shell to a column $\Sigma$ is therefore
\begin{eqnarray}
\tau(E,\Sigma) &=&
\int_0^\Sigma \kappa_{\rm bf}(E)\,d\Sigma' \nonumber \\
&\simeq& \kappa_{\rm H} \Sigma
\left(\frac{E}{\chi_{\rm H}}\right)^{-\beta}
\equiv \tau_{\rm H}(\Sigma)
\left(\frac{E}{\chi_{\rm H}}\right)^{-\beta},
\end{eqnarray}
where $ \tau_{\rm H}(\Sigma)\equiv \kappa_{\rm H}\Sigma$. The total optical depth through the full shell at the hydrogen edge is
\be
\tau_{\rm H,tot}\equiv \kappa_{\rm H}\Sigma_{\rm tot}
\simeq 4.3\times10^3\,
M_1R_{0.1}^{-2}.
\ee
Thus the shell is extremely optically thick to photons near the hydrogen
ionization threshold. However, because the opacity decreases steeply with
energy, sufficiently hard X-rays can penetrate to much larger columns. The
photon energy for which the optical depth to column $\Sigma$ is unity is
\begin{eqnarray}
E_{\tau=1}(\Sigma)
&\simeq&
\chi_{\rm H}\tau_{\rm H}^{1/\beta} \nonumber \\
&\simeq&
0.39\,{\rm keV}\,
\left(\frac{\Sigma}{\Sigma_{\rm tot}}\right)^{1/\beta}
M_1^{1/\beta}R_{0.1}^{-2/\beta},
\label{eq:Etau}
\end{eqnarray}
where the prefactor in the final expression uses $\beta=2.5$. Because $E_{\tau=1}\lesssim E_{\rm max}$, the hardest
X-rays can reach deep into the shell.

Accounting for attenuation, the local specific energy absorption rate per unit
mass at depth $\Sigma$ is
\begin{align}
&\dot{\epsilon}(\Sigma) =
 \int_{\chi_{\rm H}}^{E_{\rm max}}
\frac{1}{4\pi R^{2}}
\left(\frac{dL}{dE}\right)_0
\exp[-\tau(E,\Sigma)]
\kappa_{\rm bf}(E)\,dE \nonumber \\
&=
\frac{L_{\rm X}\kappa_{\rm H}}{4\pi R^{2}}
\frac{2-\Gamma}{E_{\rm max}^{2-\Gamma}}
\int_{\chi_{\rm H}}^{E_{\rm max}}
E^{1-\Gamma}
\left(\frac{E}{\chi_{\rm H}}\right)^{-\beta}
\exp\left[-\tau_{\rm H}(\Sigma)
\left(\frac{E}{\chi_{\rm H}}\right)^{-\beta}\right]dE
\nonumber \\
&\simeq 
1.3\frac{L_{\rm X}\kappa_{\rm H}}{4\pi R^{2}}
\left(\frac{\chi_{\rm H}}{E_{\rm max}}\right)^{1.3}
\int_{1}^{\frac{E_{\rm max}}{\chi_{\rm H}}}
y^{0.3-\beta}
\exp\left[-\tau_{\rm H}(\Sigma)y^{-\beta}\right]dy ,
\label{eq:edot}
\end{align}
where in the final line we specialize to $\Gamma=0.7$ and have substituted $y\equiv E/\chi_{\rm H}$.

Near the inner surface, where $\tau_{\rm H}\ll 1$, the absorption rate is
approximately independent of depth. This optically thin layer has a very small mass column,
$\Sigma\lesssim\kappa_{\rm H}^{-1}$, corresponding even if fully ionized to
only ${\rm DM}\lesssim 5\times10^{-2}\,{\rm pc\,cm^{-3}}$, and therefore
contributes negligibly to the total DM. Deeper in the shell, the attenuation of photons
near the hydrogen edge causes $\dot{\epsilon}$ to decrease with $\Sigma$. For columns satisfying
\be
1 \ll \tau_{\rm H}(\Sigma) \ll
\left(\frac{E_{\rm max}}{\chi_{\rm H}}\right)^\beta ,
\label{eq:intermediate}
\ee
photons near the hydrogen edge have already been absorbed, but photons with
$E\gtrsim E_{\tau=1}(\Sigma)$ continue to penetrate. For the fiducial shell, the upper condition is easily satisfied even at the
back of the shell, since
$E_{\tau=1}(\Sigma_{\rm tot}) \ll E_{\rm max}$ (Eq.~\eqref{eq:Etau}).
In this regime of greatest interest, the integral in Eq.~\eqref{eq:edot} is
dominated by photons with $y\sim \tau_{\rm H}^{1/\beta}$. Because this energy lies well above the hydrogen edge but below
$E_{\rm max}/\chi_{\rm H}$, we may extend the lower and upper limits of the
integral in Eq.~\eqref{eq:edot} to zero and infinity, respectively, yielding
\be
\int_0^\infty
y^{1-\Gamma-\beta}
\exp[-\tau_{\rm H}y^{-\beta}]\,dy
=
C_{\Gamma,\beta}\,
\tau_{\rm H}^{-\eta},
\qquad
\eta \equiv 1-\frac{2-\Gamma}{\beta}.
\ee
where
\be
\eta \equiv \frac{\beta+\Gamma-2}{\beta},
\qquad
C_{\Gamma,\beta}
\equiv
\frac{1}{\beta}
\Gamma_{\rm f}\left(\frac{\beta+\Gamma-2}{\beta}\right),
\ee
and $\Gamma_{\rm f}$ denotes the Gamma function. For our fiducial values
$\Gamma=0.7$ and $\beta=2.5$, $\eta\simeq 0.48$ and
$C_{\Gamma,\beta}\simeq 0.74$.

The absorbed X-ray energy rate per unit mass in the intermediate-depth regime is therefore
\begin{eqnarray}
&&\dot{\epsilon}(\Sigma)
\simeq
(2-\Gamma)C_{\Gamma,\beta}
\frac{L_{\rm X}\kappa_{\rm H}}{4\pi R^{2}}
\left(\frac{\chi_{\rm H}}{E_{\rm max}}\right)^{2-\Gamma}
\left[
\tau_{\rm H,tot}
\left(\frac{\Sigma}{\Sigma_{\rm tot}}\right)
\right]^{-\eta} \nonumber \\
&\approx& 8.7\times10^{3}\,{\rm erg\,g^{-1}\,s^{-1}} \times \nonumber \\
&&
L_{{\rm X},40}\,
M_1^{-0.48}
R_{0.1}^{-1.04}
\left(\frac{E_{\rm max}}{50\,{\rm keV}}\right)^{-1.3}
\left(\frac{\Sigma}{\Sigma_{\rm tot}}\right)^{-0.48}.
\label{eq:heating}
\end{eqnarray}
where $L_{{\rm X},40}\equiv L_{\rm X}/10^{40}\,{\rm erg\,s^{-1}}$ and in the second line we have again taken $\Gamma=0.7$, $\beta=2.5$.

If a fraction $f_{\rm ion}$ of $\dot{\epsilon}$ goes into
ionizing hydrogen, the ionization rate per neutral hydrogen atom is
\begin{eqnarray}
\zeta_{\rm H}(\Sigma)
&\simeq&
2.8\times10^{-10}\,{\rm s^{-1}}\,
\left(\frac{f_{\rm ion}}{0.3}\right)L_{{\rm X},40}\,
M_1^{-0.48}R_{0.1}^{-1.04} \nonumber \\
&&\times
\left(\frac{E_{\rm max}}{50\,{\rm keV}}\right)^{-1.3}
\left(\frac{\Sigma}{\Sigma_{\rm tot}}\right)^{-0.48}.
\label{eq:zetaH}
\end{eqnarray}
The parameter $f_{\rm ion}$ accounts for the fact that an absorbed X-ray
photon does not deposit all of its energy directly into ionizing hydrogen.
After the primary photoionization, the energetic photoelectron loses energy
through secondary ionizations, excitations, and Coulomb heating. Calculations
of fast-electron degradation in weakly ionized gas find that a fraction of
order a few tenths of the photoelectron energy is converted into secondary
ionizations of hydrogen \citep[e.g.,][]{ShullVanSteenberg1985,
FurlanettoStoever2010}. This fraction decreases as the gas becomes more highly
ionized, because Coulomb losses to thermal electrons increasingly dominate.
Since we work in the weakly ionized limit, we hereafter adopt $f_{\rm ion}=0.3$ as a
fiducial value and absorb it into the numerical normalizations below.\footnote{In what follows, the
ionization rates scale linearly with $f_{\rm ion}$, while the equilibrium
ionization fractions and DMs scale as $f_{\rm ion}^{1/2}$ and the
recombination timescale scales as $f_{\rm ion}^{-1/2}$.}

Equation~\eqref{eq:zetaH} also gives the instantaneous growth rate of the
ionized column when recombinations are negligible. In this early-time limit,
$dx/dt\simeq\zeta_{\rm H}$, so the cumulative DM grows at a rate
\be
\dot{\rm DM}_{\rm ion}(<\Sigma)
\simeq
\frac{X}{m_p}\int_0^\Sigma \zeta_{\rm H}(\Sigma')\,d\Sigma' .
\ee
Using $\zeta_{\rm H}\propto(\Sigma/\Sigma_{\rm tot})^{-\eta}$ in the
intermediate-depth regime gives
\be
\dot{\rm DM}_{\rm ion}(<\Sigma)
\simeq
\frac{{\rm DM}_{\rm max}\zeta_{\rm H}(\Sigma_{\rm tot})}
{1-\eta}
\left(\frac{\Sigma}{\Sigma_{\rm tot}}\right)^{1-\eta}.
\label{eq:DMdot_ion_cumulative}
\ee
Thus the maximum photoionization-driven growth rate through the shell is
\begin{eqnarray}
&&\dot{\rm DM}_{\rm ion,tot}
\simeq
\frac{{\rm DM}_{\rm max}\zeta_{\rm H}(\Sigma_{\rm tot})}
{1-\eta} \nonumber \\
&&\simeq
3.8\,{\rm pc\,cm^{-3}\,yr^{-1}}\,
L_{{\rm X},40}
M_1^{0.52}
R_{0.1}^{-3.04}
\left(\frac{E_{\rm max}}{50\,{\rm keV}}\right)^{-1.3}.
\label{eq:DMdot_ion_tot}
\end{eqnarray}
This expression provides a direct connection between the ionizing luminosity
and the initial DM growth rate, before recombinations become important.

Photoionization is opposed by radiative recombination. In the weakly ionized
limit, $n_{\rm H}\simeq X\rho/m_p$ and $n_e\simeq n_p\simeq xn_{\rm H}$, so
the recombination rate per unit volume is
\be
\dot{n}_{\rm rec}
=
\alpha_{\rm rec}(T)n_e n_p
\simeq
\alpha_{\rm rec}(T)x^2 n_{\rm H}^2 .
\label{eq:nrec}
\ee
For hydrogen case-B recombination we adopt
\be
\alpha_{\rm rec}(T)
\simeq
2.6\times10^{-13}\,{\rm cm^3\,s^{-1}}\,
T_4^{-0.7},
\ee
where $T_4\equiv T/10^4\,{\rm K}$. We assume that the gas swept into the shell
has had time to cool radiatively (Appendix~\ref{sec:sn-shell}), so that the
temperature entering this expression is set by thermal balance between
photoelectric heating (Eq.~\eqref{eq:heating}) and radiative cooling. In Appendix~\ref{sec:thermal-balance}, we show that this balance places the
gas near the low-temperature edge of the atomic cooling curve, somewhat below
$10^{4}\,{\rm K}$.

\subsection{Equilibrium Ionization and Dispersion Measure}

We next estimate the equilibrium ionization fraction by balancing
photoionization against radiative recombination.  At the equilibrium temperatures expected in the dense shell, obtained by balancing photoelectric heating against radiative cooling (Appendix~\ref{sec:thermal-balance}), collisional ionization can be neglected relative to photoionization. In collisional ionization equilibrium, the electron
fraction falls to negligible values just below $T\sim10^4\,{\rm K}$
(\citealt{Schure2009}, their Table~2), whereas the ionized fraction in our
model is maintained by the external X-ray flux.

 The photo-ionization rate per unit volume is
$\zeta_{\rm H}n_{\rm H}(1-x)$, which in the weakly ionized limit becomes
$\simeq\zeta_{\rm H}n_{\rm H}$. Equating this to Eq.~\eqref{eq:nrec} gives
\be
\zeta_{\rm H} n_{\rm H}
\simeq
\alpha_{\rm rec}(T)x_{\rm eq}^2 n_{\rm H}^2 ,
\label{eq:ion_rec_balance}
\ee
or
\be
x_{\rm eq}(\Sigma)
\simeq
\left[
\frac{\zeta_{\rm H}(\Sigma)}
{\alpha_{\rm rec}(T)n_{\rm H}}
\right]^{1/2}.
\label{eq:xion_rec}
\ee
Using the ionization rate estimated above (Eq.~\eqref{eq:zetaH}), this gives
\begin{eqnarray}
x_{\rm eq}(\Sigma)
&\simeq&
7.0\times10^{-2}\,
L_{{\rm X},40}^{1/2}
M_1^{-0.74}
R_{0.1}^{0.98}
\left(\frac{f_\Delta}{0.01}\right)^{1/2} \nonumber \\
&&\times
\left(\frac{E_{\rm max}}{50\,{\rm keV}}\right)^{-0.65}
\left(\frac{\Sigma}{\Sigma_{\rm tot}}\right)^{-0.24}
T_4^{0.35}.
\label{eq:xion_num}
\end{eqnarray}
For fiducial parameters, $x_{\rm eq}\ll1$, justifying the weakly ionized
approximation adopted above.

The cumulative equilibrium DM interior to column $\Sigma$ is
\be
{\rm DM}_{\rm eq}(<\Sigma)
=
\frac{X}{m_p}
\int_0^\Sigma x_{\rm eq}(\Sigma')\,d\Sigma' .
\ee
Since $x_{\rm eq}\propto(\Sigma/\Sigma_{\rm tot})^{-\eta/2}$,
\be
{\rm DM}_{\rm eq}(<\Sigma)
=
{\rm DM}_{\rm eq,tot}
\left(\frac{\Sigma}{\Sigma_{\rm tot}}\right)^{1-\eta/2},
\ee
where
\be
{\rm DM}_{\rm eq,tot}
\simeq
\frac{{\rm DM}_{\rm max}x_{\rm eq}(\Sigma_{\rm tot})}
{1-\eta/2}.
\label{eq:DMeq_cumulative}
\ee
For $\Gamma=0.7$ and $\beta=2.5$, this gives
\be
\frac{{\rm DM}_{\rm eq}(<\Sigma)}
{{\rm DM}_{\rm eq,tot}}
\simeq
\left(\frac{\Sigma}{\Sigma_{\rm tot}}\right)^{0.76},
\ee
with
\begin{eqnarray}
{\rm DM}_{\rm eq,tot}
&\simeq&
20\,{\rm pc\,cm^{-3}}\,
L_{{\rm X},40}^{1/2}
M_1^{0.26}
R_{0.1}^{-1.02}
\left(\frac{f_\Delta}{0.01}\right)^{1/2} \nonumber \\
&&\times
\left(\frac{E_{\rm max}}{50\,{\rm keV}}\right)^{-0.65}
T_4^{0.35}.
\label{eq:DMeq_tot}
\end{eqnarray}
for the total DM through the shell in ionization equilibrium.   Although the ionization fraction is largest near the inner edge, the equilibrium DM is weighted toward deeper layers of the shell as long as $1-\eta/2>0$, which is generically satisfied.

The scalings in Eq.~\eqref{eq:DMeq_tot} have simple physical interpretations. A higher
X-ray luminosity increases the photoionization rate and therefore raises the
equilibrium ionized fraction, giving ${\rm DM}_{\rm eq,tot}\propto L_{\rm X}^{1/2}$.
Increasing the shell mass increases the total available column, but it also
raises the density and optical depth, which enhances recombination and shifts
the absorbed radiation to higher photon energies; the net result is the weak
scaling ${\rm DM}_{\rm eq,tot}\propto M_1^{0.26}$. A smaller shell radius both
increases the incident X-ray flux and increases the gas density and column, but
the latter effects dominate slightly, giving ${\rm DM}_{\rm eq,tot}\propto R_{0.1}^{-1.02}$. Finally, increasing
$E_{\rm max}$ at fixed $L_{\rm X}$ hardens the spectrum by placing more of the
energy in photons with smaller bound-free opacity, reducing the absorbed power
and hence ${\rm DM}_{\rm eq,tot}\propto E_{\rm max}^{-0.65}$.

The timescale to approach this ionization-recombination equilibrium is of
order the recombination time evaluated at the equilibrium electron density,
\be
t_{\rm rec,eq}(\Sigma)
\simeq
\frac{1}{\alpha_{\rm rec}n_e}
=
\frac{1}{\alpha_{\rm rec}x_{\rm eq}n_{\rm H}} .
\ee
Using Eq.~\eqref{eq:xion_num}, this gives
\begin{eqnarray}
t_{\rm rec,eq}(\Sigma)
&\simeq&
7.5\,{\rm yr}\,
L_{{\rm X},40}^{-1/2}
M_1^{-0.26}
R_{0.1}^{2.02}
\left(\frac{f_\Delta}{0.01}\right)^{1/2} \nonumber \\
&&\times
\left(\frac{E_{\rm max}}{50\,{\rm keV}}\right)^{0.65}
\left(\frac{\Sigma}{\Sigma_{\rm tot}}\right)^{0.24}
T_4^{0.35}.
\label{eq:trec}
\end{eqnarray}
Thus the inner, more strongly ionized layers equilibrate somewhat faster,
while the outer layers equilibrate more slowly. However, because the depth
dependence is weak, $t_{\rm rec}\propto(\Sigma/\Sigma_{\rm tot})^{0.24}$, the
equilibration time remains of order a few years across most of the shell for
fiducial parameters.  


\subsection{Time-Evolving X-ray Luminosity}
\label{sec:time-dependent}

Thus far we have assumed a time-steady ionizing luminosity, so that each layer
of the shell relaxes toward the equilibrium ionization fraction derived above.
In reality, the X-ray luminosity of the central engine will likely vary with time, for
example due to enhanced activity of a magnetar central engine. We now consider how the shell DM responds to finite episodes of enhanced activity.

At a fixed depth $\Sigma$, the time-dependent ionization fraction obeys
\be
\frac{dx}{dt}
=
\zeta_{\rm H}(\Sigma,t)(1-x)
-
\alpha_{\rm rec}n_{\rm H}x^2 .
\ee
In the weakly ionized limit, $1-x\simeq1$, this can be written
\be
\frac{dx}{dt}
=
\zeta_{\rm H}(\Sigma,t)
\left[
1-\left(\frac{x}{x_{\rm eq}(\Sigma,t)}\right)^2
\right],
\label{eq:xdot}
\ee
where $x_{\rm eq}(\Sigma,t)$ is the instantaneous equilibrium value defined by
Eq.~\eqref{eq:xion_rec}, evaluated using the current ionizing luminosity.

The associated recombination time is
\be
t_{\rm rec,eq}(\Sigma)
\equiv
\frac{x_{\rm eq}}{\zeta_{\rm H}}
=
\frac{1}{\alpha_{\rm rec}n_{\rm H}x_{\rm eq}} .
\label{eq:trec_def}
\ee

As an example, consider a finite episode in which $L_{\rm X}$ is constant for
a duration $t_{\rm active}$ and negligible before and after. Starting from neutral gas, $x(0)=0$, photoionization initially increases the
ionized fraction linearly, before recombinations become important and the gas
approaches equilibrium according to
\be
x(t,\Sigma)
=
x_{\rm eq}(\Sigma)
\tanh\left[
\frac{t}{t_{\rm rec,eq}(\Sigma)}
\right],
\qquad
0<t<t_{\rm active}.
\label{eq:x_on_solution}
\ee
Thus $x\simeq\zeta_{\rm H}t$ at early times and
$x\rightarrow x_{\rm eq}$ after a few $t_{\rm rec,eq}$. At the end of the active phase,
\be
x \equiv x_{\rm active}(\Sigma) =
x_{\rm eq}(\Sigma)
\tanh\left[
\frac{t_{\rm active}}{t_{\rm rec,eq}(\Sigma)}
\right].
\ee
After the luminosity turns off, the gas recombines,
$dx/dt=-\alpha_{\rm rec}n_{\rm H}x^2$, giving
\begin{eqnarray}
&& x(t,\Sigma)
= \nonumber \\
&&
\frac{x_{\rm active}(\Sigma)}
{1+
\left[x_{\rm active}(\Sigma)/x_{\rm eq}(\Sigma)\right]
(t-t_{\rm active})/t_{\rm rec,eq}(\Sigma)} ,
\qquad \nonumber \\
t>t_{\rm active},
\label{eq:x_off_solution}
\end{eqnarray}
Short activity episodes therefore produce an approximately linear DM rise
followed by a slower recombination-powered decline, while long episodes allow
the shell to approach ${\rm DM}_{\rm eq,tot}$ before the ionizing luminosity
fades.

\begin{figure}
\centering
\includegraphics[width=0.48\textwidth]{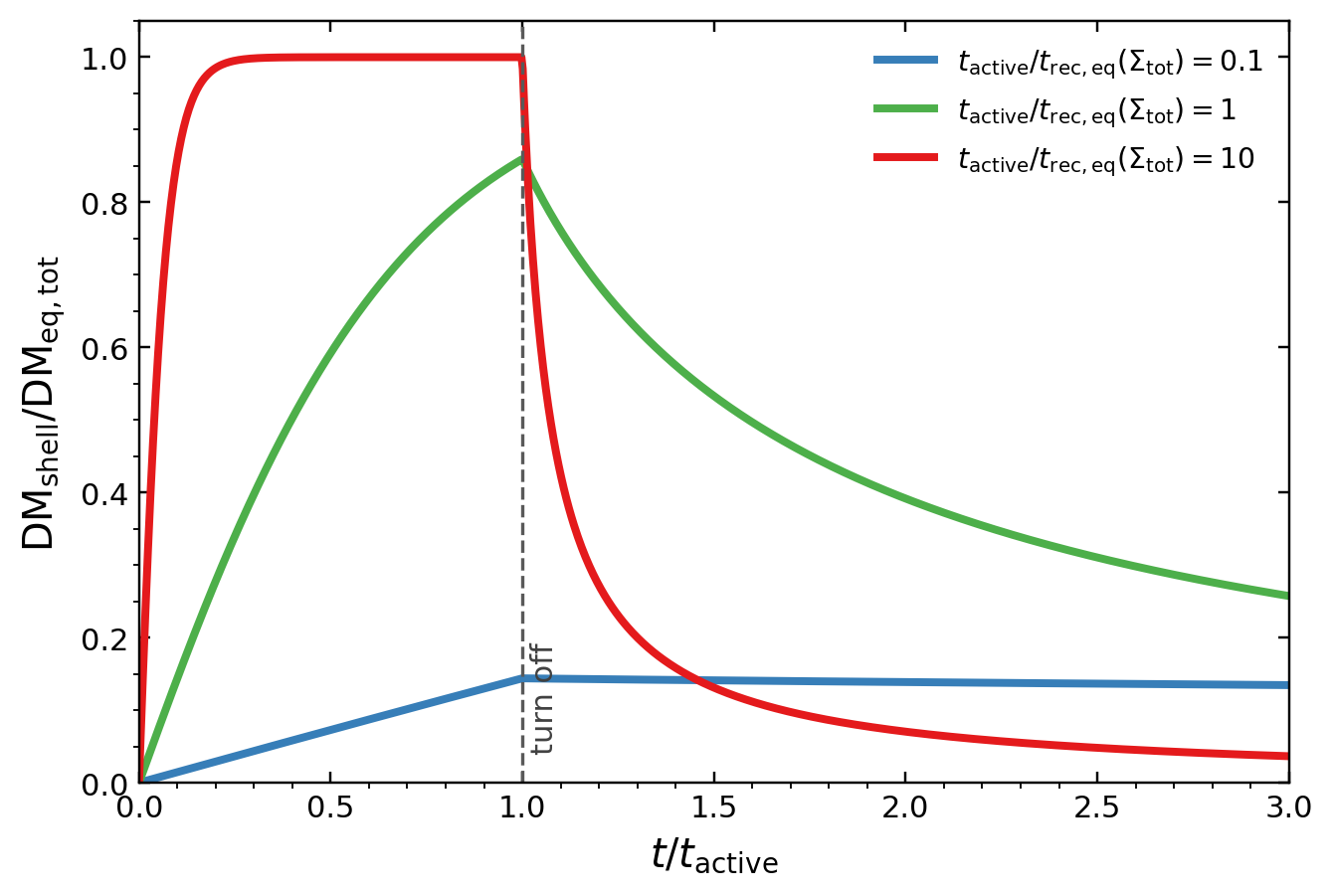}
\includegraphics[width=0.48\textwidth]{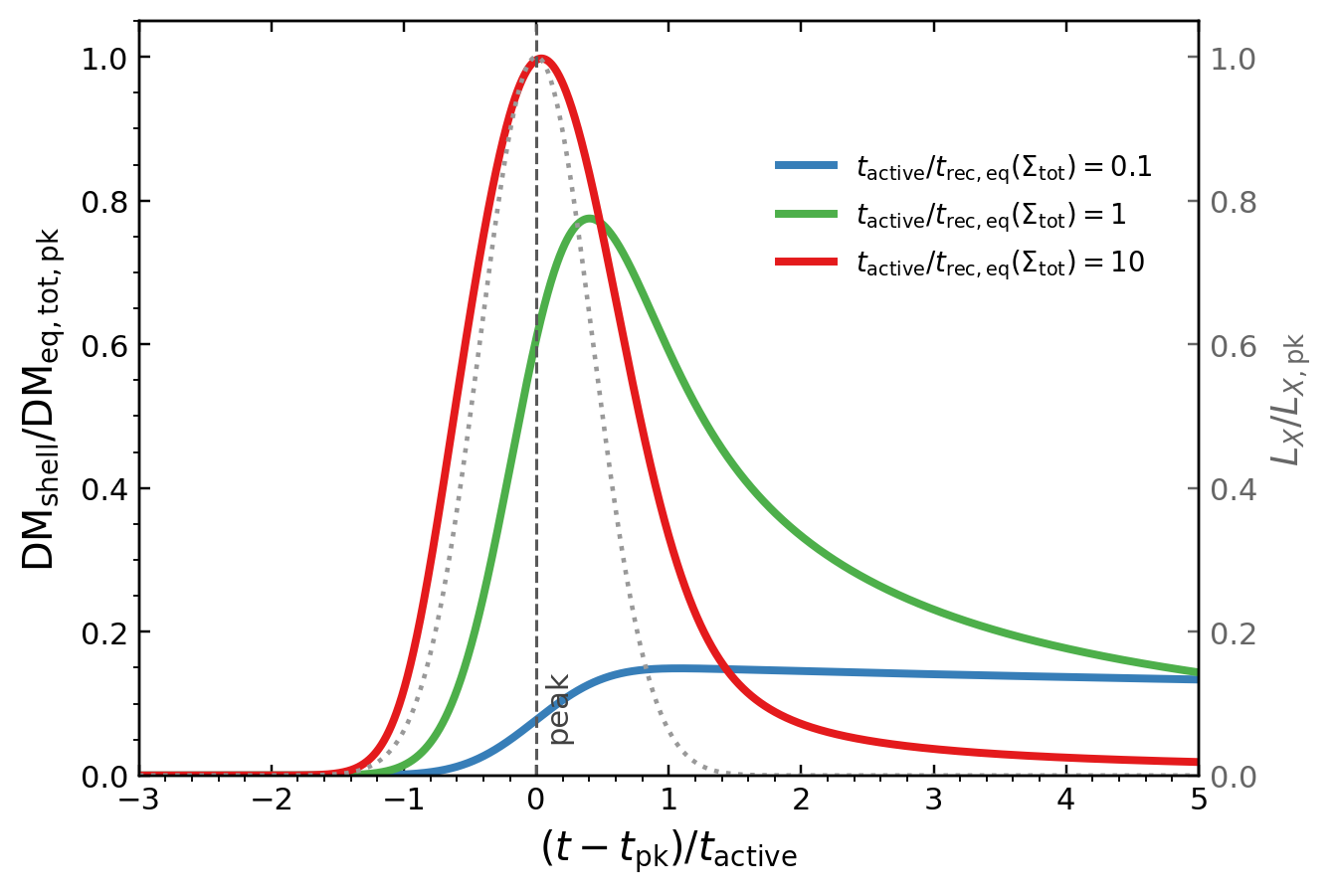}
\caption{Evolution of the DM response during and after finite episodes of ionizing
X-ray activity. The upper panel shows the analytic solution for an idealized
constant-luminosity episode of duration $t_{\rm active}$
(Eqs.~\ref{eq:x_on_solution} and \ref{eq:x_off_solution}); the lower panel
shows the result of numerically solving Eq.~\eqref{eq:xdot} for a Gaussian
luminosity history with full-width half-maximum $t_{\rm active}$. Curves are
labeled by $t_{\rm active}/t_{\rm rec,eq}(\Sigma_{\rm tot})$ and are
normalized to the equilibrium DM corresponding to a steady luminosity equal to
the peak luminosity of the episode. The curves may be interpreted as the
response of a representative layer near $\Sigma\simeq\Sigma_{\rm tot}$, which
is a good proxy for the shell-integrated DM because both the equilibrium DM and
the initial DM growth rate are weighted toward large columns. For slowly
varying luminosity, the DM remains close to equilibrium and tracks
${\rm DM}_{\rm eq}\propto L_{\rm X}^{1/2}$. For short activity episodes, the
DM rises approximately with the accumulated ionizing fluence, reaches only a
fraction of ${\rm DM}_{\rm eq,tot}$, and then undergoes a slower
recombination-powered decline after the luminosity fades.
}
\label{fig:DM_response}
\end{figure}

More general luminosity histories produce qualitatively similar behavior, with
$t_{\rm active}$ replaced by the timescale over which $L_{\rm X}$ changes
appreciably. Figure~\ref{fig:DM_response} illustrates the shell response for
two idealized finite episodes of enhanced ionizing activity. The upper panel shows the analytic solution for the constant-luminosity episode
described above, while the lower panel shows the result of solving
Eq.~\eqref{eq:xdot} numerically for a Gaussian luminosity history with
full-width half-maximum $t_{\rm active}$. In both cases, the curves are
normalized to the equilibrium DM corresponding to a steady luminosity equal to
the peak luminosity of the episode.

The behavior depends primarily on the ratio
$t_{\rm active}/t_{\rm rec,eq}(\Sigma_{\rm tot})$. If the luminosity varies
slowly compared to the recombination time, the shell remains close to
ionization equilibrium and the DM approximately tracks the instantaneous
equilibrium value, ${\rm DM}_{\rm eq}\propto L_{\rm X}^{1/2}$. Conversely, for
short activity episodes, the DM does not reach its equilibrium value. It rises
approximately in proportion to the accumulated ionizing fluence, but after the
luminosity declines the gas recombines more slowly because the ionized fraction
never became large.

\section{Discussion}
\label{sec:discussion}

\subsection{Supernova Remnant Shells}
\label{sec:remnant}

Up to this point, our treatment has been largely agnostic as to the origin of
the radius, mass, and thickness of the dense shell. However, the most natural
application is to supernova ejecta swept up by the expanding magnetar wind
nebula.

As discussed in Appendix~\ref{sec:sn-shell}, for supernova ejecta of kinetic
energy $E_{\rm SN}=10^{51}E_{51}\,{\rm erg}$ and total mass
$M_{\rm ej}=10M_{\rm ej,10}M_{\odot}$, the nebular radius, corresponding to
the radius of the swept-up shell, is
\be
R=R_{\rm n}
\simeq
7.3\times10^{-2}\,{\rm pc}\,
E_{{\rm mag},50}^{1/5}
E_{51}^{3/10}
M_{\rm ej,10}^{-1/2}
\left(\frac{t_{\rm age}}{50\,{\rm yr}}\right),
\label{eq:Rn_disc}
\ee
where $E_{\rm mag}=10^{50}E_{{\rm mag},50}\,{\rm erg}$ is the total energy
injected into the nebula. The corresponding swept-up shell mass is
\be
M=M_{\rm sh}
\simeq
0.49\,M_{\odot}\,
M_{\rm ej,10}
E_{{\rm mag},50}^{3/5}
E_{51}^{-3/5}.
\label{eq:Msh_disc}
\ee
The remaining shell parameter is its fractional thickness,
$f_\Delta=\Delta/R_{\rm n}$.  Gas swept through the nebular forward shock cools rapidly compared to the
system age for the densities of interest (Appendix~\ref{sec:sn-shell}), while
the photoionized shell temperature is regulated near $T\sim10^4\,{\rm K}$ and
provides little thermal pressure support against compression
(Appendix~\ref{sec:thermal-balance}). If the cooled shell were supported only
by this thermal pressure, very large compression factors,
$\chi_{\rm max}\sim10^2$--$10^3$, would be possible, corresponding to a
one-dimensional cooling thickness as small as
$\Delta/R_{\rm n}\sim10^{-3}$ (Eq.~\ref{eq:Delta_min}). In reality, the
effective thickness is likely set by the nonlinear structure of the cooled
shell, including thin-shell/thermal/Rayleigh-Taylor instabilities and associated turbulence. These effects can broaden the shell to $f_\Delta=\Delta/R_{\rm n}\sim10^{-2}$, motivating the fiducial value adopted above (see Eq.~\ref{eq:Delta_inst}).

We now specialize the ionization calculation of Section~\ref{sec:model} to
this supernova-remnant shell. Substituting Eqs.~\eqref{eq:Rn_disc} and
\eqref{eq:Msh_disc} into Eq.~\eqref{eq:DMeq_tot}, we find
\begin{eqnarray}
&&{\rm DM}_{\rm eq,tot}
\simeq
23\,{\rm pc\,cm^{-3}}\,
L_{{\rm X},40}^{1/2}
M_{\rm ej,10}^{0.77}
E_{{\rm mag},50}^{-0.05}
E_{51}^{-0.46} \times \nonumber \\
&& 
\left(\frac{t_{\rm age}}{50\,{\rm yr}}\right)^{-1.02} \left(\frac{f_\Delta}{0.01}\right)^{1/2}
\left(\frac{E_{\rm max}}{50\,{\rm keV}}\right)^{-0.65}
T_4^{0.35}
\, .
\label{eq:DMeq_SN}
\end{eqnarray}
The corresponding maximum photoionization-driven DM growth rate, applicable
before recombinations become important, follows from Eq.~\eqref{eq:DMdot_ion_tot}:
\begin{eqnarray}
\dot{\rm DM}_{\rm ion,tot}
&\simeq&
6.8\,{\rm pc\,cm^{-3}\,yr^{-1}}\,
L_{{\rm X},40}
M_{\rm ej,10}^{2.04}
E_{{\rm mag},50}^{-0.30}
E_{51}^{-1.22} \nonumber \\
&&\times
\left(\frac{t_{\rm age}}{50\,{\rm yr}}\right)^{-3.04}
\left(\frac{E_{\rm max}}{50\,{\rm keV}}\right)^{-1.3}.
\label{eq:DMdot_SN}
\end{eqnarray}
Finally, the recombination time at the outer edge of the shell is
\begin{eqnarray}
t_{\rm rec,eq}(\Sigma_{\rm tot})
&\simeq&
4.8\,{\rm yr}\,
L_{{\rm X},40}^{-1/2}
M_{\rm ej,10}^{-1.27}
E_{{\rm mag},50}^{0.25}
E_{51}^{0.76}
\left(\frac{t_{\rm age}}{50\,{\rm yr}}\right)^{2.02} \nonumber \\
&&\times
\left(\frac{f_\Delta}{0.01}\right)^{1/2}
\left(\frac{E_{\rm max}}{50\,{\rm keV}}\right)^{0.65}
T_4^{0.35}.
\label{eq:trec_SN}
\end{eqnarray}
Equating $t_{\rm rec,eq}$ to the age of the system suggests that ionization
equilibrium, $t_{\rm rec,eq}\lesssim t_{\rm age}$, can be maintained for a few
hundred years for fiducial parameters, after which the local shell contribution to the DM will have declined to only a few ${\rm pc\,cm^{-3}}$. One can also verify from
Eq.~\eqref{eq:xion_num} that the weak-ionization assumption,
$x_{\rm eq}\ll1$, remains valid over this period.

These estimates imply two conceptually distinct sources of DM evolution, as illustrated in the right panel of Fig.~\ref{fig:cartoon}. The
first is the secular evolution of the shell itself as the nebula expands into
the homologous ejecta. This effect operates on the age of the system $\sim t_{\rm age}$ and is
therefore closely related to earlier models in which expanding supernova ejecta
or remnants produce a declining local DM. The physical interpretation is
different, however. Here the measured DM is not the full ionized ejecta column,
but the ionized fraction selected by photoionization-recombination balance. As
the shell expands, its density decreases and the equilibrium ionized fraction
rises, but the declining column still wins. At fixed ionizing luminosity and fixed total injected magnetar energy,
Eq.~\eqref{eq:DMeq_SN} gives a secular decline
\be
{\rm DM}_{\rm eq,tot}\propto t_{\rm age}^{-\delta_{\rm DM}},
\qquad
\delta_{\rm DM}
=
\frac{3}{2} -\eta
=
\frac{1}{2} +\frac{2-\Gamma}{\beta}.
\label{eq:delta_DM}
\ee
For the fiducial values
$\Gamma=0.7$ and $\beta=2.5$, this gives
$\delta_{\rm DM}\simeq1.02$. More generally, for
$\Gamma\simeq0.5$--$1.5$, broadly motivated by fits to the lower-energy part
of magnetar-burst spectra, one obtains
$\delta_{\rm DM}\simeq 0.7$--$1.1$ for $\beta=2.5$. This is shallower than
the usual ${\rm DM}\propto t^{-2}$ scaling expected for a freely expanding
shell with a fixed ionized fraction \citep[e.g.,][]{Connor2016}.  

Another physically distinct secular effect is possible if the unshocked
ejecta ahead of the nebular forward shock are already ionized. In that case,
radiative cooling behind the shock can reduce the free-electron column by
sweeping ionized gas into the cold, recombined shell. Appendix~\ref{sec:sn-shell}
shows that this effect can contribute a negative drift of order
$\dot{\rm DM}_{\rm sh}\sim-{\rm few}\,{\rm pc\,cm^{-3}\,yr^{-1}}$ for the
fiducial parameters. Thus shock processing and engine photoionization need not
act in the same direction: the former can provide a slowly declining baseline,
while enhanced engine irradiation can temporarily raise the ionized column of
the dense shell.

The second source of variability is a change in the ionizing luminosity $L_{\rm X}$ at approximately fixed shell structure (Sec.~\ref{sec:time-dependent}), i.e. on a timescale $t_{\rm active} \ll t_{\rm age}$.
Such changes can drive faster DM fluctuations on the recombination time rather
than on the system age. Following an increase in $L_{\rm X}$, the DM can
initially rise at a rate as large as Eq.~\eqref{eq:DMdot_SN}, before
recombinations become important and the shell approaches the new ionization
equilibrium. When the ionizing luminosity decreases, the shell recombines on
the timescale in Eq.~\eqref{eq:trec_SN}, producing the delayed and asymmetric
DM response illustrated schematically in Fig.~\ref{fig:cartoon} and by the examples in Fig.~\ref{fig:DM_response}.

Portions of the ejecta exterior to the swept-up shell, at radii $r>R_{\rm n}$, can also
be photoionized and contribute to the source-local DM. However, this extended
component responds much more slowly than the swept-up shell because of its lower density and longer recombination time. To see this, one
can regard the unswept ejecta as an effective shell at a characteristic radius
$R_{\rm t} \gtrsim R_{\rm n}$, with fractional thickness $f_\Delta\sim1$ and mass of
order $M_{\rm ej}$. From Eq.~\eqref{eq:trec}, the recombination time scales as
$t_{\rm rec,eq}\propto M^{-0.26}R^{2.02}f_\Delta^{1/2}$. Relative to the thin
swept-up shell, the larger radius and much larger radial thickness of the
unswept ejecta more than compensate for its larger mass, giving a response
time longer by $1-2$ orders of magnitude. Any DM contribution from the extended
ejecta is therefore expected to behave as an approximately constant local DM
on the year-to-decade timescales of interest here.

\subsection{Application to FRB~121102}
\label{sec:FRB121102}

\begin{figure*}
\centering
\includegraphics[width=1.5\columnwidth]{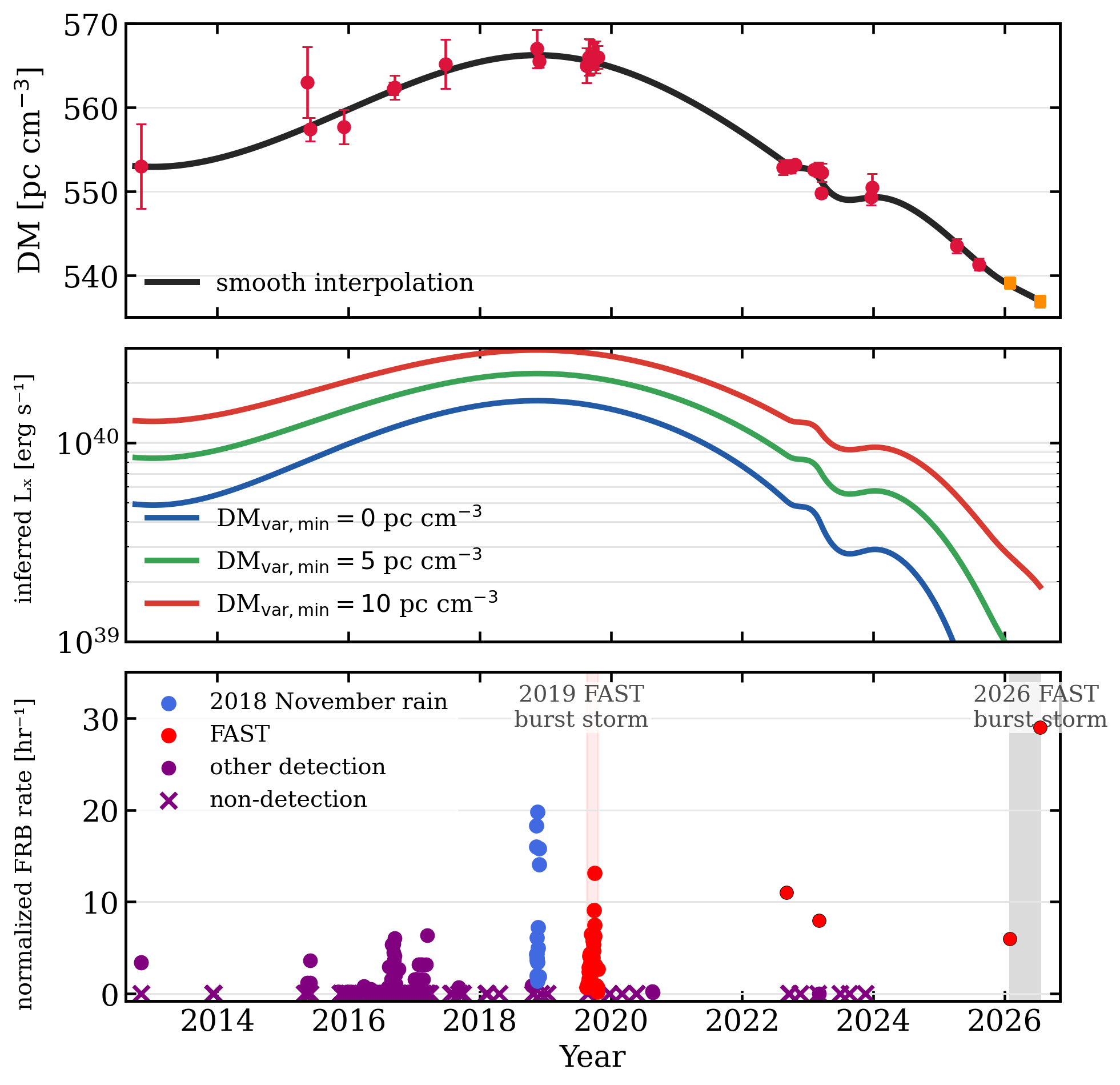}
\caption{
Comparison of the observed DM evolution of FRB~121102 with the ionizing
luminosity history inferred in the photoionization model, and with the radio
burst activity history. Top: binned DM measurements from Table~2 of
\citet{Waxman2026}, together with recent FAST measurements reported by
\citet{ATel17921,ATel17642}. The black curve shows a smooth interpolation used
only to guide the eye and to estimate the slowly varying component of the DM.
Middle: ionizing luminosity inferred from the interpolated DM evolution using
Eq.~\eqref{eq:LX_from_DM}, shown for three choices of the minimum variable
source-local contribution, ${\rm DM}_{\rm var,min}$. Bottom: normalized
L-band FRB activity rates, following the fluence-threshold normalization of
\citet{Braga2025}. Purple circles and crosses denote detections and
non-detections, respectively; blue points mark the 2018 November ``rain'' of
bursts; red points mark FAST detections, including the 2019 FAST burst storm
and the more recent high-activity FAST epochs reported by \citet{Wang2025,ATel17642,ATel17921}. The 2026 FAST rates have been
rescaled to the same fiducial fluence threshold used for the normalized
activity rates. The shaded regions mark the 2019 and 2026 FAST burst-storm
epochs.
}
\label{fig:FRB121102}
\end{figure*}

FRB~121102 provides the best current test case for this model because it has
both a compact persistent radio source \citep{Chatterjee2017,Marcote2017} and
a measured secular evolution in DM \citep{Hessels2019,Wang2025,Snelders2025,
Waxman2026}. Recent long-baseline monitoring and compilations show that the
DM increased over the decade leading up to a maximum around 2019--2020 and has
declined by $\sim 25\,{\rm pc\,cm^{-3}}$ in the subsequent years (see top panel of Fig.~\ref{fig:FRB121102}). The approximately comparable rise
and decay times suggest that the variable component was not responding in the
strongly impulsive limit. Instead, the ionized shell was likely close to
photoionization-recombination equilibrium during much of the evolution, as in
the slowly varying examples shown in Fig.~\ref{fig:DM_response}.

We interpret the rising phase as an interval of enhanced time-averaged
ionizing luminosity from the central engine, plausibly associated with an epoch of increased burst activity, and the declining phase as the subsequent reduction of this luminosity followed by recombination of the dense
shell. In this limit the observed variable DM approximately traces the
instantaneous equilibrium value,
\be
{\rm DM}_{\rm var}(t)
\simeq
{\rm DM}_{\rm eq,tot}[L_{\rm X}(t)] ,
\ee
up to an additive contribution from more extended material surrounding the engine that is approximately constant on decade timescales. Because ${\rm DM}_{\rm eq,tot}\propto L_{\rm X}^{1/2}$, the ionizing
luminosity required to produce a variable shell DM is
\begin{eqnarray}
L_{\rm X}
&\simeq&
10^{40}\,{\rm erg\,s^{-1}}\,
\left(
\frac{{\rm DM}_{\rm var}}
{23\,{\rm pc\,cm^{-3}}}
\right)^2
M_{\rm ej,10}^{-1.54}
E_{{\rm mag},50}^{0.10}
E_{51}^{0.92}
\nonumber \\
&&\times
\left(\frac{t_{\rm age}}{50\,{\rm yr}}\right)^{2.04} \left(\frac{f_\Delta}{0.01}\right)^{-1}
\left(\frac{E_{\rm max}}{50\,{\rm keV}}\right)^{1.3}
T_4^{-0.7}.
\label{eq:LX_from_DM}
\end{eqnarray}
This estimate is within an order of magnitude of the ionizing luminosity
expected if a modest fraction of the long-term magnetar power needed to inflate
the nebula emerges in hard X-rays (Eq.~\ref{eq:LX_engine}).

The corresponding recombination time follows by substituting
Eq.~\eqref{eq:LX_from_DM} into Eq.~\eqref{eq:trec_SN}. At the outer edge of
the shell this gives
\begin{eqnarray}
t_{\rm rec,eq}(\Sigma_{\rm tot})
&\simeq&
4.8\,{\rm yr}\,
\left(
\frac{{\rm DM}_{\rm var}}
{23\,{\rm pc\,cm^{-3}}}
\right)^{-1}
M_{\rm ej,10}^{-1/2}
E_{{\rm mag},50}^{0.20}
\nonumber \\
&&\times
E_{51}^{0.30} 
\left(\frac{t_{\rm age}}{50\,{\rm yr}}\right)
\left(\frac{f_\Delta}{0.01}\right)
T_4^{0.70}.
\label{eq:trec_FRB121102}
\end{eqnarray}
Thus, for fiducial parameters, the shell can approach ionization equilibrium
on a timescale of order years, short enough to track the observed decade-scale
rise and decline of the variable DM. 

A useful way to compare the model with the observed DM evolution is to invert
the measured DM into the ionizing luminosity required by the shell. This
exercise is necessarily uncertain because the observed DM contains an unknown
constant contribution from the Milky Way, intergalactic medium, host galaxy,
and slowly responding local material. We therefore write
${\rm DM}_{\rm obs}(t)={\rm DM}_{\rm const}+{\rm DM}_{\rm var}(t)$ and infer
$L_{\rm X}(t)$ from Eq.~\eqref{eq:LX_from_DM}, adopting the fiducial shell
parameters used there. Figure~\ref{fig:FRB121102} shows the result for several
choices of the residual variable shell DM at the minimum observed DM epoch.
These choices shift the normalization of the inferred luminosity, while leaving
its qualitative time dependence unchanged: $L_{\rm X}$ peaks near the epoch of
maximum DM and remains within the energetically plausible range estimated
above.

The bottom panel of Fig.~\ref{fig:FRB121102} compares this inferred ionizing
luminosity history with a heterogeneous proxy for the radio activity, following
Fig.~3 of \citet{Braga2025}.
The timing is suggestive because the strongest documented radio activity
occurred shortly before the DM maximum. In the published L-band sample, the
November 2018 Arecibo burst storm represents the highest absolute
detection-rate epoch \citep{Jahns2023,Braga2025}; FRB~121102 was also detected
by CHIME/FRB during this same active interval \citep{Josephy2019}. The subsequent
2019 FAST campaign detected thousands of bursts \citep{Li2021Nature},
confirming that the source remained in an unusually active state near the
epoch of maximum DM. In the photoionization picture, this sequence is natural if the radio activity
traces the time-averaged ionizing luminosity: enhanced engine activity raises
the equilibrium ionized column, while the later DM decline reflects the
subsequent fading of that ionizing luminosity. 

We caution that comparison is necessarily qualitative. The radio activity is not a direct measurement of the ionizing luminosity, and the published rates combine
heterogeneous observing campaigns with different fluence thresholds, observing
bands, and selection effects. Nevertheless, the strongest radio activity
episodes occur close to the epoch over which the model requires enhanced
$L_{\rm X}$, supporting the possibility that the DM maximum was driven by
heightened engine activity.

The most recent activity history should therefore be especially informative.
FAST monitoring indicates that FRB~121102 was relatively quiet in 2020--2021,
reactivated in 2022--2023, and has recently entered another highly active
state \citep{Wang2025,ATel17642,ATel17921}. The 2026 FAST observations are
particularly constraining because they report high burst rates while the DM
continues to decline, with ${\rm DM}=539.1\pm0.1\,{\rm pc\,cm^{-3}}$ in late
January and ${\rm DM}=536.9\pm0.1\,{\rm pc\,cm^{-3}}$ in mid-July
\citep{ATel17642,ATel17921}. In the present model, the implication depends on
whether the renewed radio activity is accompanied by a comparable increase in
the time-averaged ionizing luminosity. If it is, the DM decline should flatten and eventually reverse; if not, the
continued decline would indicate that the radio burst rate is not a reliable
proxy for the ionizing luminosity, or that another component dominates the
measured DM evolution.

\section{Conclusions and Predictions}
\label{sec:conclusions}

We have shown that time-dependent X-ray photoionization can produce DM
variations of order tens of ${\rm pc\,cm^{-3}}$ on year-to-decade timescales
in young magnetar environments. In this picture, the observed local DM does not
measure the total gas column through the ejecta, but only the ionized column
maintained by the variable high-energy output of the central engine. The
required ionizing luminosity is a modest fraction of the long-term engine power
needed to inflate the observed persistent radio sources, provided that active
extragalactic repeaters partition their flare energy between radio and X-rays
in a manner broadly similar to the Galactic magnetar burst FRB~200428.
Although such X-ray luminosities are difficult to detect directly for most
extragalactic repeaters, except perhaps for the very nearest sources, they can
still regulate the ionization state of the surrounding ejecta. The evolving DM
may therefore provide an unusually sensitive indirect probe of the
time-averaged high-energy output of the central engine.

The photoionization scenario makes several observational predictions:
\begin{enumerate}
\item \textit{Long-term activity should correlate with source-local DM.}
Epochs of enhanced burst activity should be accompanied by increases in the
source-local DM, provided that the radio activity traces the time-averaged
ionizing luminosity. The response need not be simultaneous: if the luminosity varies on a timescale
comparable to or shorter than the recombination time, the DM should lag the
activity and decay more slowly after the activity subsides (Fig.~\ref{fig:DM_response}). The relevant timescale is not the recombination
time of the innermost ionized skin, but the value near the large-column layers
that dominate the shell-integrated DM, typically of order years for the
FRB~121102 parameters considered above. The renewed FAST activity of FRB~121102 in 2026 is already probing this
prediction: despite burst rates comparable to, or exceeding, previous active
epochs, the reported DM values remain on the declining branch
\citep{ATel17642,ATel17921}. If the radio activity traces a renewed increase
in the time-averaged ionizing luminosity, the model predicts that the DM
decline should flatten and may reverse over the coming years; continued decline
would instead imply that the radio burst rate is not a reliable proxy for the
ionizing luminosity, or that another component dominates the measured DM
evolution.

\item \textit{DM and RM variability need not track one another in detail.}
In this picture, the variable DM is produced mainly in the cool, dense shell,
where the free-electron column responds to changes in the ionizing luminosity
through photoionization and recombination. By contrast, the large RM is
produced primarily in the hotter magnetized nebula, where changes can reflect
magnetic-field geometry, turbulence, expansion, and baryon loading. Enhanced
engine activity may therefore influence both observables, but a one-to-one
short-term correlation between DM and RM is not expected.

\item \textit{Energetic bursts should produce small prompt DM increases.}
Consider a radio burst of isotropic-equivalent energy $E_{\rm r}$ accompanied
by X-ray energy $E_{\rm X}=\eta_{\rm X/r}E_{\rm r}$, released on a timescale
short compared to the recombination time. In the limit that the induced change
in ionization is small, recombination can be neglected during the burst.
Replacing $L_{\rm X}dt$ in Eq.~\eqref{eq:DMdot_SN} by $E_{\rm X}$ gives
\begin{eqnarray}
&&\Delta{\rm DM}_{\rm ion,tot}
\simeq
2.1\times10^{-2}\,{\rm pc\,cm^{-3}}\,
\left(\frac{\eta_{\rm X/r}}{10^{5}}\right)
\left(\frac{E_{\rm r}}{10^{40}\,{\rm erg}}\right)
 \nonumber \\
&&\times
M_{\rm ej,10}^{2.04} E_{{\rm mag},50}^{-0.30}
E_{51}^{-1.22}
\left(\frac{t_{\rm age}}{50\,{\rm yr}}\right)^{-3.04}
\left(\frac{E_{\rm max}}{50\,{\rm keV}}\right)^{-1.3}
.
\label{eq:DeltaDM_burst}
\end{eqnarray}
Individual bursts with $E_{\rm r}\gtrsim10^{40}\,{\rm erg}$ can in principle
produce DM jumps approaching the precision of the best propagation
measurements, especially for younger, denser shells, softer ionizing spectra,
or larger X-ray-to-radio fluence ratios. In practice, however, detecting such
steps after single bursts will be difficult because FRB bursts often show
complex frequency-dependent morphology, including sub-burst drift, unresolved
components, scattering, and intrinsic spectral structure, all of which can bias
the DM value that maximizes burst sharpness or time-frequency structure
\citep[e.g.,][]{Hessels2019,Josephy2019,Li2021Nature}. The cleaner
observable is therefore likely to be a statistically coherent DM increase
across many bursts following an especially energetic burst or burst-rich
episode, measured with a consistent propagation-DM estimator.

\item \textit{The secular DM decline should be shallower than $t^{-2}$.}
On timescales comparable to the age of the system, the shell DM should decline
as the ejecta expand. However, this decline need not follow the naive
${\rm DM}\propto t^{-2}$ scaling of a freely expanding shell with fixed
ionization fraction. In the photoionization-regulated case, expansion lowers
the density and increases the equilibrium ionized fraction, partially
offsetting the declining column. For the swept-up supernova shell considered
above, this gives
${\rm DM}_{\rm eq,tot}\propto t_{\rm age}^{-\delta_{\rm DM}}$, with $\delta_{\rm DM}\simeq 0.7$--$1.1$ for plausible magnetar-burst spectral slopes. This exponent describes the response at fixed ionizing luminosity; if instead $L_{\rm X}$ declines secularly with the engine power, the DM decline would be
correspondingly steeper.

\end{enumerate}

Applied to FRB~121102, these results place its DM evolution within the same
``concordance'' picture previously developed for this source: a hyper-active
young magnetar, born in an otherwise ordinary core-collapse supernova, inflates
a compact baryon-loaded nebula whose size, synchrotron luminosity,
electron-ion composition, and large but secularly declining RM are set by the
long-term magnetar energy input and confinement by the surrounding ejecta
\citep{Metzger2017,MargalitMetzger2018,Beloborodov2017,Metzger2019}. In this interpretation, the time-variable DM arises from the ionized fraction
of that same ejecta shell, regulated by the variable X-ray output of the
central engine. This differs from models in which the DM evolution more directly
traces the expansion of a PRS-driven shocked shell \citep{Waxman2026}; however,
shock interaction and engine photoionization need not act in the same
direction, since sweeping already ionized upstream ejecta into the largely
neutral shell can itself contribute a negative DM drift
(Eq.~\eqref{eq:DMdot_shock_recomb}).

Several simplifications in the present calculation should be revisited with
more detailed photoionization models. We have followed primarily the ionization
balance of hydrogen and adopted an approximate bound-free opacity law
appropriate to solar-composition material. Supernova ejecta can be substantially metal-enriched relative to solar
composition. At the modest ionization parameters relevant
here, only the outer, valence electrons of heavier elements may be efficiently
photoionized, but the enhanced metal abundance will increase the bound-free
opacity and hence the fraction of incident X-ray radiation absorbed by the
shell. We therefore expect composition-dependent photoionization calculations
to change the quantitative mapping between $L_{\rm X}$ and DM, while leaving
the basic mechanism of ionization-regulated DM variability intact.

Although we have focused on supernova ejecta shells irradiated by flaring
magnetar engines, the basic mechanism is more general. The calculation in
Section~\ref{sec:model} is agnostic to the origin of the dense shell: any FRB
engine embedded behind a partially neutral, recombining column and producing
time-variable ionizing radiation could generate qualitatively similar DM
variability. One possible example is a mass-transferring X-ray binary, whose
black-hole or neutron-star jet has been proposed as an FRB engine capable of
explaining binary-periodic activity \citep{Sridhar2021}. In such systems,
super-Eddington disk winds can inflate wind bubbles bounded by dense shells,
which may produce persistent radio nebulae and local propagation effects
qualitatively similar to those discussed here \citep{Sridhar2022}.

Continued monitoring that combines precise, consistently measured DMs with
radio activity histories and high-energy constraints will test whether FRB
dispersion measures are tracing not only the amount of plasma near the source,
but the time-dependent ionizing output of the central engine.

\acknowledgments

I am grateful to Laura Spitler and Ani Patel for helpful discussions.  I thank Tony Piro for carefully reading an early version of the manuscript and Crist\'obal Braga Vi\~nals for providing information used to construct the FRB activity plot in Fig.~\ref{fig:FRB121102}.  The Flatiron Institute is supported by the Simons Foundation.

\appendix

\section{Equilibrium Temperature of the Photoionized Shell}
\label{sec:thermal-balance}

As discussed in Appendix~\ref{sec:sn-shell}, gas heated by the shock driven
into the supernova ejecta by the nebula cools radiatively on a timescale
shorter than the system age, motivating the dense radiative shell adopted in our
calculations. The equilibrium temperature of the irradiated shell is then set
by the competition between photoelectric heating
(Section~\ref{sec:model}) and radiative cooling. If a fraction $f_{\rm heat}$
of the absorbed X-ray power is ultimately converted into heat, the volumetric
heating rate at depth $\Sigma$ is
\be
\dot{q}_{\rm heat}
\simeq
f_{\rm heat}\rho\dot{\epsilon}.
\ee
Using Eq.~\eqref{eq:edot}, this gives the heating rate of gas at depth $\Sigma$ through the shell,
\begin{eqnarray}
\dot{q}_{\rm heat}
&\simeq&
4.8\times10^{-15}\,{\rm erg\,cm^{-3}\,s^{-1}}\,
f_{\rm heat}\,
L_{{\rm X},40}\,
M_1^{0.52}
R_{0.1}^{-4.04}
\left(\frac{f_\Delta}{0.01}\right)^{-1} 
\left(\frac{E_{\rm max}}{50\,{\rm keV}}\right)^{-1.3}
\left(\frac{\Sigma}{\Sigma_{\rm tot}}\right)^{-0.48}.
\end{eqnarray}
We expect $f_{\rm heat}\sim1-f_{\rm ion}$ and hence of order unity. Although
the primary photoelectron energy is shared between secondary ionization,
excitation, and Coulomb heating, the UV and soft-X-ray photons produced by
excitation are likely to be reabsorbed in the neutral shell, converting much of
this energy into heat or additional ionization.

For gas near solar composition, metal-line cooling dominates over free-free
cooling for $T\sim10^4-10^6\,{\rm K}$, with a cooling coefficient
$\Lambda(T)\sim{\rm few}\times10^{-22}\,{\rm erg\,cm^3\,s^{-1}}$ near the
line-cooling peak \citep[e.g.,][]{Schure2009}. The corresponding cooling rate is
\be
\dot{q}_{\rm cool}
\simeq
n_e n_{\rm H}\Lambda(T)
\simeq
x n_{\rm H}^2\Lambda(T).
\ee
Evaluated at the ionization-recombination equilibrium value
$x=x_{\rm eq}(\Sigma)$ from Eq.~\eqref{eq:xion_num}, this becomes
\begin{eqnarray}
\dot{q}_{\rm cool}
&\simeq&
3.7\times10^{-13}\,{\rm erg\,cm^{-3}\,s^{-1}}\,
\Lambda_{-22}\,
L_{{\rm X},40}^{1/2}
M_1^{1.26}
R_{0.1}^{-5.02}
\left(\frac{f_\Delta}{0.01}\right)^{-3/2} 
\left(\frac{E_{\rm max}}{50\,{\rm keV}}\right)^{-0.65}
\left(\frac{\Sigma}{\Sigma_{\rm tot}}\right)^{-0.24}
T_4^{0.35},
\end{eqnarray}
where $\Lambda_{-22}\equiv\Lambda/10^{-22}\,{\rm erg\,cm^3\,s^{-1}}$.  
Thermal balance ($\dot{q}_{\rm heat} = \dot{q}_{\rm cool}$) therefore requires
\begin{eqnarray}
\Lambda(T)
&\simeq&
1.3\times10^{-24}\,{\rm erg\,cm^3\,s^{-1}}\,
\frac{f_{\rm heat}}{f_{{\rm ion},0.3}^{1/2}}
L_{{\rm X},40}^{1/2}
M_1^{-0.74}
R_{0.1}^{0.98}
\left(\frac{f_\Delta}{0.01}\right)^{1/2} 
\left(\frac{E_{\rm max}}{50\,{\rm keV}}\right)^{-0.65}
\left(\frac{\Sigma}{\Sigma_{\rm tot}}\right)^{-0.24}
T_4^{-0.35}.
\label{eq:Lambda_req}
\end{eqnarray}
This required value is well below the peak metal-line cooling coefficient,
$\Lambda\sim10^{-22}$--$10^{-21}\,{\rm erg\,cm^3\,s^{-1}}$, for
solar-metallicity gas at $T\sim10^5$--$10^6\,{\rm K}$
\citep[e.g.,][]{Schure2009}. The gas therefore cools below the peak of the
cooling curve until the cooling efficiency drops rapidly near the
low-temperature edge of the atomic cooling curve. The tabulated cooling
coefficients of \citet{Schure2009} provide a useful calibration: their
coefficient normalized to $n_en_{\rm H}$, $\Lambda_N$, falls through the
fiducial value required by Eq.~\eqref{eq:Lambda_req},
$\Lambda\sim{\rm few}\times10^{-25}\,{\rm erg\,cm^3\,s^{-1}}$, at
$\log T\simeq3.85$--$3.9$, corresponding to
$T\simeq(7$--$8)\times10^3\,{\rm K}$ (their Table~2). At the same temperatures,
the collisional-ionization-equilibrium electron fraction in their final column
is only $n_e/n_{\rm H}\sim10^{-4}$, well below the photoionization fractions of
interest here. Thus thermal balance naturally places the gas near
$T\lesssim10^4\,{\rm K}$, while the free-electron fraction remains controlled
primarily by photoionization and recombination.

\section{Origin of the Shell: Supernova Ejecta}
\label{sec:sn-shell}

\subsection{Ejecta Structure and Nebular Expansion}

The primary motivation for the dense shell considered in this work is the shocked and cooled ejecta swept up by the expanding magnetar nebula. We consider supernova ejecta of total mass $M_{\rm ej}=10M_{\rm ej,10}M_\odot$ and kinetic energy $E_{\rm SN}=10^{51}E_{51}\,{\rm erg}$. At times much longer than
the explosion dynamical time, the ejecta expand homologously, such that
each fluid element moves with constant velocity $v$ and has radius
$r=vt$. We adopt a standard broken power-law density profile
\citep[e.g.][]{Chevalier1989,MatznerMcKee1999},
\begin{equation}
\rho_{\rm ej}(v,t)=
\frac{\zeta_\rho M_{\rm ej}}
{v_t^3t^3}
\left\{
\begin{array}{ll}
(v/v_t)^{-\delta}, & v<v_t, \\[6pt]
(v/v_t)^{-n}, & v>v_t,
\end{array}
\right.
\label{eq:rhoej}
\end{equation}
where
\begin{equation}
\zeta_\rho=
\frac{(3-\delta)(n-3)}
{4\pi(n-\delta)}.
\end{equation}
The inner and outer ejecta are separated by the transition velocity
\begin{equation}
v_t
=
\left[
\frac{2(5-\delta)(n-5)}
{(3-\delta)(n-3)}
\frac{E_{\rm SN}}
{M_{\rm ej}}
\right]^{1/2}.
\end{equation}
For the fiducial choice $\delta=0$ and $n=10$, this gives
$\zeta_\rho=21/(40\pi) \simeq0.17$ and
\begin{equation}
v_t \simeq
3.5\times10^3\,{\rm km\,s^{-1}}\,
E_{51}^{1/2}
M_{\rm ej,10}^{-1/2}.
\label{eq:vt_delta0}
\end{equation}

We are primarily interested in the inner ejecta layers encountered by the
expanding magnetar wind nebula. For $\delta=0$, these layers have
$v<v_t$ and possess an approximately constant density,
\begin{equation}
\rho_{\rm ej,t}
\simeq
2.0\times10^{-20}\,{\rm g\,cm^{-3}}\,
M_{\rm ej,10}^{5/2}
E_{51}^{-3/2}
\left(\frac{t}{50\,{\rm yr}}\right)^{-3}.
\label{eq:rhot_delta0}
\end{equation}

The central engine injects a magnetized outflow into the expanding ejecta with
long-term average power $\dot E_{\rm mag}\sim E_{\rm mag}/t_{\rm age}$, the
same quantity used in Section~\ref{sec:model} to motivate the fiducial
ionizing luminosity. This outflow inflates a hot nebula bounded by a forward
shock that sweeps the inner ejecta into a dense shell. During the early phase,
before the supernova reverse shock returns to the center and while the magnetar
power is approximately steady, the expansion of this pulsar/magnetar wind
nebula is described by the standard self-similar solution
\citep{Chevalier1977,Chevalier1984,GaenslerSlane2006}. Evaluated at the system
age $t_{\rm age}$, the forward-shock radius is
\begin{eqnarray}
R_{\rm n}
&\simeq&
7.3\times10^{-2}\,{\rm pc}\,
E_{{\rm mag},50}^{1/5}
E_{51}^{3/10}
M_{\rm ej,10}^{-1/2} \left(\frac{t_{\rm age}}{50\,{\rm yr}}\right),
\label{eq:Rn_age}
\end{eqnarray}
where $E_{\rm mag}=10^{50}E_{{\rm mag},50}\,{\rm erg}$ is the total energy
injected by the magnetar. This is comparable to the characteristic shell radius
adopted in Section~\ref{sec:model} for ages of a few decades to a century.

The transition radius between the inner flat ejecta core and outer steep
envelope is $R_t=v_t t_{\rm age}$. Using Eqs.~\eqref{eq:vt_delta0} and
\eqref{eq:Rn_age}, the nebula remains within the inner ejecta core for
fiducial parameters,
\be
\frac{R_{\rm n}}{R_t}\simeq0.41\,
E_{{\rm mag},50}^{1/5}
E_{51}^{-1/5}.
\ee
The mass swept into the dense shell is approximately the ejecta mass originally
interior to the nebular forward shock,
\be
M_{\rm sh}
\simeq
\int_0^{R_{\rm n}}4\pi r^2\rho_{\rm ej,t}\,dr
=
\frac{4\pi}{3}\rho_{\rm ej,t}R_{\rm n}^3 \simeq
\frac{4\pi}{3}\zeta_\rho M_{\rm ej}
\left(\frac{R_{\rm n}}{R_t}\right)^3,
\ee
where in the final line we have used $\rho_{\rm ej,t}=\zeta_\rho M_{\rm ej}/(v_t^3t_{\rm age}^3)$. For $\delta=0$, $n=10$, $\zeta_\rho=21/(40\pi)$, this becomes
\be
M_{\rm sh}\simeq
0.49\,M_\odot\,
M_{\rm ej,10}
E_{{\rm mag},50}^{3/5}
E_{51}^{-3/5},
\ee
where in the final equality we have used Eq.~\eqref{eq:Rn_age}.

\subsection{Radiative Cooling and Shell Thickness}

We next check whether the swept-up material can cool efficiently enough to form
the thin dense shell assumed in the main text. The relevant heating occurs at
the forward shock driven by the expanding nebula into the freely expanding
ejecta. The velocity relevant for shock heating is therefore the velocity of
the nebular forward shock relative to the upstream ejecta. 
Since the unshocked ejecta at radius $R_{\rm n}$ move with velocity
$R_{\rm n}/t$, we have
\be
v_{\rm sh}
=
\frac{dR_{\rm n}}{dt}
-
\frac{R_{\rm n}}{t}
=
\frac{1}{5}\frac{R_{\rm n}}{t},
\ee
where in the final equality we used the self-similar
scaling $R_{\rm n}\propto t^{6/5}$, before substituting
$\dot E_{\rm mag}\sim E_{\rm mag}/t_{\rm age}$ to obtain
Eq.~\eqref{eq:Rn_age}. Using Eq.~\eqref{eq:Rn_age}, evaluated at $t=t_{\rm age}$, we find
\be
v_{\rm sh}
\simeq
290\,{\rm km\,s^{-1}}\,
E_{{\rm mag},50}^{1/5}
E_{51}^{3/10}
M_{\rm ej,10}^{-1/2}.
\label{eq:vsh}
\ee
The immediate post-shock temperature and density follow from the strong-shock
jump conditions,
\begin{eqnarray}
T_{\rm sh}
&\simeq&
\frac{3}{16}
\frac{\mu m_p}{k_{\rm B}}
v_{\rm sh}^2 \simeq 1.1\times10^6\,{\rm K}\,
\left(\frac{\mu}{0.62}\right)
E_{{\rm mag},50}^{2/5}
E_{51}^{3/5}
M_{\rm ej,10}^{-1},
\end{eqnarray}
and
\be
\rho_{\rm ps}
\simeq
4\rho_{\rm ej,t}
\simeq
8.2\times10^{-20}\,{\rm g\,cm^{-3}}\,
M_{\rm ej,10}^{5/2}
E_{51}^{-3/2}
\left(\frac{t_{\rm age}}{50\,{\rm yr}}\right)^{-3},
\ee
The radiative cooling time of the shocked gas is
\be
t_{\rm cool}
\simeq
0.4\,{\rm yr}\,
\left(\frac{T_{\rm sh}}{1.1\times10^6\,{\rm K}}\right)
\Lambda_{-21}^{-1}
M_{\rm ej,10}^{-5/2}
E_{51}^{3/2}
\left(\frac{t_{\rm age}}{50\,{\rm yr}}\right)^3.
\ee
where $\Lambda_{-21}\equiv
\Lambda/(10^{-21}\,{\rm erg\,cm^3\,s^{-1}})$, and we have assumed a fully
ionized plasma of approximately solar composition. Around
$T\sim10^5$--$10^6\,{\rm K}$, solar-metallicity cooling curves reach
$\Lambda\sim10^{-22}$--$10^{-21}\,{\rm erg\,cm^3\,s^{-1}}$
\citep[e.g.,][]{Schure2009}, so the shocked gas cools much faster than the
system age for fiducial parameters.

If $t_{\rm cool}\lesssim t_{\rm age}$, the shocked gas cools radiatively after
passing through the forward shock, from $T_{\rm sh}$ down to the
low-temperature end of the atomic cooling curve, which we denote by
$T_{\rm min}\sim10^4\,{\rm K}$. If the cooling occurs approximately at
constant pressure $P \propto \rho T$, the total compression relative to the upstream ejecta density is
\begin{eqnarray}
\chi_{\rm max}
&\equiv&
\frac{\rho_{\rm cool}}{\rho_{\rm ej,t}}
\simeq
4\frac{T_{\rm sh}}{T_{\rm min}} \simeq
440\,
\left(\frac{T_{\rm min}}{10^4\,{\rm K}}\right)^{-1}
\left(\frac{\mu}{0.62}\right)
E_{{\rm mag},50}^{2/5}
E_{51}^{3/5}
M_{\rm ej,10}^{-1},
\end{eqnarray}
where the factor of 4 is the immediate post-shock compression. The cooled
swept-up material is therefore geometrically thin. If the swept-up mass is
concentrated near $R_{\rm n}$ in a shell of density
$\rho_{\rm cool}=\chi_{\rm max}\rho_{\rm ej,t}$, its characteristic thickness is
\be
\Delta_{\rm min}
\simeq
\frac{M_{\rm sh}}{4\pi R_{\rm n}^{2}\rho_{\rm cool}}.
\ee
Using $M_{\rm sh}=(4\pi/3)\rho_{\rm ej,t} R_{\rm n}^{3}$ for the flat inner ejecta,
this gives
\begin{eqnarray}
\frac{\Delta_{\rm min}}{R_{\rm n}}
&\simeq&
\frac{1}{3\chi_{\rm max}} \simeq
7.6\times10^{-4}\,
\left(\frac{T_{\rm min}}{10^4\,{\rm K}}\right)
\left(\frac{\mu}{0.62}\right)^{-1}
E_{{\rm mag},50}^{-2/5}
E_{51}^{-3/5}
M_{\rm ej,10}.
\label{eq:Delta_min}
\end{eqnarray}
This should be regarded as the minimum thickness expected from a
one-dimensional, isobarically cooling flow. In reality, radiatively cooled
shells are subject to thin-shell instabilities \citep[e.g.,][]{Vishniac1994},
as well as Rayleigh-Taylor instabilities, turbulent mixing, and magnetic
pressure support. These effects can broaden the shell and prevent it from
reaching the extreme compression implied by pure 1D cooling.

A rough lower limit on the effective shell thickness is set by the shortest
wavelength on which the cooled shell can respond hydrodynamically over the
system age. Taking this scale to be of order $c_{\rm s}t_{\rm age}$, where
$c_{\rm s}$ is the sound speed of the cooled gas, gives
\be
\frac{\Delta}{R_{\rm n}}
\gtrsim
\frac{c_{\rm s}t_{\rm age}}{R_{\rm n}}
\simeq
{\cal M}^{-1},
\ee
where ${\cal M}\equiv R_{\rm n}/(c_{\rm s}t_{\rm age})$ is the effective Mach
number of the shell expansion relative to gas at $T_{\rm min}$. Numerically,
\begin{eqnarray}
\frac{\Delta}{R_{\rm n}}
&\gtrsim&
8.1\times10^{-3}\,
\left(\frac{T_{\rm min}}{10^4\,{\rm K}}\right)^{1/2}
\left(\frac{\mu_{\rm cool}}{0.62}\right)^{-1/2} 
E_{{\rm mag},50}^{-1/5}
E_{51}^{-3/10}
M_{\rm ej,10}^{1/2},
\label{eq:Delta_inst}
\end{eqnarray}
where we used $c_{\rm s}=(k_{\rm B}T_{\rm min}/\mu_{\rm cool}m_p)^{1/2}$ and
Eq.~\eqref{eq:Rn_age}. Thus, although radiative cooling alone would imply an
extremely thin shell, multidimensional effects plausibly regulate the effective
thickness to values of order $\Delta/R_{\rm n}\sim10^{-2}$, comparable to the
fiducial value adopted in the main text.

Thus, although radiative cooling alone could produce compression factors of
hundreds, corresponding to
$f_\Delta\sim\Delta_{\rm min}/R_{\rm n}\sim10^{-3}$, multidimensional
instabilities, turbulent mixing, and magnetic pressure support may limit the
effective compression to $\chi_{\rm eff}\sim10$--$100$. This implies
$f_\Delta\simeq(3\chi_{\rm eff})^{-1}\sim3\times10^{-3}$--$3\times10^{-2}$,
motivating the fiducial value $f_\Delta=10^{-2}$ adopted in the main text
while emphasizing its uncertainty.

This radiative cooling also implies that the forward shock need not increase
the observed DM. If the freely expanding ejecta upstream of the shock are
already highly ionized, sweeping them into the cold dense shell removes ionized
column at a rate
\be
\dot{\rm DM}_{\rm sh}
\simeq
-\frac{X\rho_{\rm ej,t}v_{\rm sh}}{m_p}
\simeq
-2.5\,{\rm pc\,cm^{-3}\,yr^{-1}}\,
E_{{\rm mag},50}^{1/5}
M_{\rm ej,10}^{2}
E_{51}^{-6/5}
\left(\frac{t_{\rm age}}{50\,{\rm yr}}\right)^{-3},
\label{eq:DMdot_shock_recomb}
\ee
where we have kept only the hydrogen contribution to the electron column. Thus, in the radiative-shock limit, shock processing of already ionized ejecta can produce a negative DM drift of order a few ${\rm pc\,cm^{-3}\,yr^{-1}}$ for the fiducial parameters, rather
than a positive DM contribution.

\subsection{Shock-Powered Luminosity}

Finally, we estimate the radiative luminosity powered by the nebular forward
shock itself. This provides a floor to the ionizing luminosity of the shell,
even in the absence of magnetar flaring, although the emergent spectrum may be
reprocessed to UV/optical wavelengths if the shocked gas is optically thick.

The rate at which freely expanding ejecta are swept through the forward shock is
\be
\dot{M}_{\rm sh}
\simeq
4\pi R_{\rm n}^2\rho_{\rm ej,t} v_{\rm sh},
\ee
where $v_{\rm sh}$ is the shock velocity relative to the upstream homologous
ejecta, given by Eq.~\eqref{eq:vsh}. If the shock is radiative, the kinetic
energy flux entering the shock is radiated with high efficiency, giving
\be
L_{\rm sh}
\simeq
\frac{1}{2}\dot{M}_{\rm sh}v_{\rm sh}^2
\simeq
2\pi R_{\rm n}^2\rho_{\rm ej,t} v_{\rm sh}^3
=
\frac{2\pi}{125}\rho_{\rm ej,t}\frac{R_{\rm n}^5}{t^3}.
\label{eq:Lsh_def}
\ee
Using Eqs.~\eqref{eq:rhot_delta0} and \eqref{eq:Rn_age}, evaluated at
$t=t_{\rm age}$, this becomes
\be
L_{\rm sh}
\simeq
1.5\times10^{38}\,{\rm erg\,s^{-1}}\,
E_{{\rm mag},50}
\left(\frac{t_{\rm age}}{50\,{\rm yr}}\right)^{-1}.
\label{eq:Lsh}
\ee
Equivalently,
\be
\frac{L_{\rm sh}}{\dot E_{\rm mag}}
\simeq
2.4\times10^{-3}.
\label{eq:Lsh_ratio}
\ee
The simplicity of this result is a consequence of the self-similar expansion:
the same ejecta density scale that determines the nebular radius also sets the
mass flux through the forward shock. As a result, the shock power is a fixed
dimensionless fraction of the engine power, with the small coefficient mainly
set by the fact that the shock overtakes the homologous ejecta only at
$v_{\rm sh}=R_{\rm n}/5t$.

\bibliographystyle{aasjournal}


\end{document}